\documentclass[10pt]{article}

\usepackage[english]{babel}
\usepackage{latexsym,amsmath,amssymb,amsbsy,amstext,amscd,amsfonts}
\usepackage{graphics,graphicx}
\usepackage{color}
\usepackage{tabularx}
\usepackage{multirow}
\usepackage{booktabs}
\usepackage{epstopdf}
\usepackage{natbib}
\usepackage{ulem}
\usepackage [latin1]{inputenc}

\date{}

\title{On the influence of fluid rheology on  hydraulic fracture }

\author{M. Wrobel$^{(1)}\footnote{Corresponding author: wrobel.michal@ucy.ac.cy}$ , G. Mishuris$^{(2)}$ and P. Papanastasiou$^{(1)}$ 
\\
{\it $^{(1)}$\! Department of Civil and Environmental Engineering, University of Cyprus, }
\\ {\it 1678 Nicosia, Cyprus}
\\
{\it $^{(2)}$Department of Mathematics, Aberystwyth University,}
\\ {\it Ceredigion SY23 3BZ, Wales, UK}
 }

\begin{document}

\maketitle

\begin{abstract}
We analyse a problem of a hydraulic fracture driven by a non-Newtonian shear-thinning fluid. Fluid viscosity is described by the four-parameter truncated power-law model. By varying the parameters of the rheological model we investigate spatial and temporal evolution of fluid flow inside the crack and the resulting fracture geometry. A detailed quantitative and qualitative analysis of the underlying physical phenomena is delivered. The results  demonstrate that rheological properties of  fluids significantly affect the process of hydraulic fracture not only by the values of viscosity, but also by the range of fluid shear rates over which variation of viscosity occurs.
\end{abstract}

\providecommand{\keywords}[1]
{
  \small	
  \textbf{\textit{Keywords:}} #1
}

\section{Introduction}
Hydraulic fracturing (HF) is a phenomenon of a hydraulically propelled crack encountered in many natural and man-made processes. One of its most prominent  applications in technology is fracking - a technique used to stimulate tight hydrocarbon reservoirs. 

The physical behaviour of the fracturing fluid is very important for successful execution of HF treatments.  In fact, it is one of very few elements of the process that can be engineered. Fluid viscosity is one of the major parameters related to the fracture design. High viscosity increases the fracture width facilitating transport of proppant and reduces the fluid loss to formation. However, its overestimation increases the cost of treatments, may cause excessive growth of the fracture height due to the treating pressure rise and finally can reduce the fracture conductivity.  On the other hand, low viscosity of fluid results in narrow fracture width. Consequently, in order to retain the desired concentration of proppant the fluid has to be pumped at high volumetric rates, which  significantly increases the load on the pumping equipment and reduces its lifetime. To prevent such undesired effects and to enhance the proppant transport and placement the so-called hybrid treatments are implemented \citep{Li_2016}. Within this approach different fluids (of differing rheological properties) are employed during subsequent stages of the fracking operation. For example, in the initial step the slickwater is injected to create a network of fractures. Such a network is more complex than that produced with crosslinked fluids \citep{Chong_2010}. Next, linear and/or crosslinked fluids of much better ability for proppant transport and placement are pumped. Finally, the encapsulated breaker can be activated at the fracture closure to produce linear fluid during the flowback. Fracturing fluids are designed so as to achieve an optimal combination of mechanical and chemical properties at a reasonable cost  \citep{Barbati_2016,Osiptsov_2017}. Moreover, the properties of fracturing fluid are also very important for the post-fracturing stage of the treatments where the fluid rheology  has a major impact on the efficiency of the fracture cleanup \citep{Osiptsov_2020}.

Most of the complex fracturing fluids exhibit non-Newtonian behaviour. In the case of shear-thinning fluids the viscosity decreases with shear rate growth. However, with real fluids this trend holds only over some range of shear rates as neither infinite (for low shear rates) or zero viscosity (for high shear rates) can be achieved in practice. Thus,  for low and high shear rates, Newtonian plateaus are reached. The high shear rate plateau corresponds to the viscosity of the base solvent \citep{Lecampion_2018}. Nevertheless, it is still not well understood how this complex characteristics affect the propagation of hydraulic fractures.

Optimal selection of the fracturing fluid and proper description of its physical properties becomes one of the most important issues in the HF design. The apparent viscosity of fluid is established in the rheometric measurements. A crucial requirement here is that testing should be performed at shear rate values representative of those encountered in the real fracture \citep{Montgomery_2013}. However, depending on the type of treatments (related e.g. to the properties of the rock formation) respective values can differ by orders of magnitude. Therefore, mathematical modelling of the HF process can constitute a powerful tool in prediction of the expected behaviour of the fracture providing valuable feedback for the measurements. 

The simplest mathematical model capable of describing the non-Newtonian properties of fracturing fluid is the power-law model \citep{Bird_1987}. Being convenient in computational implementation, the concept of power-law fluid has been frequently employed in numerical modelling of hydraulic fractures \citep{Adachi_2002,Garagash_2006,Perkowska_2016,Peck_2018_1,Peck_2018_2}. However, the simple power-law rheology cannot correctly describe the complex characteristics of a real fracturing fluid with viscosity plateaus.  Here, more advanced four-parameter models such as the ones of Carreau or the Cross are more physically adequate \citep{Bird_1987,Habibpour_2017}. On the other hand, these models are cumbersome in numerical implementation as they do not allow analytical integration of the respective flow equations to obtain the average fluid flow rate. This problem was circumvented in the work of \cite{Wrobel_Arxiv} where an efficient algorithm to model the flow of generalized Newtonian fluids in  channels of simple geometries was introduced. The proposed subroutine enables one to compute the average velocities and fluid flow rates with much better efficiency than other schemes described in the literature. Thus, it is well suited for the hydraulic fracture problems, where multiple evaluations of the fluid flow rates are performed in the computational process.

Another way to account for the complex fluid rheology is to use a simplified model which retains the most important features of the original law, while simultaneously allowing analytical integration of the flow equations. Such a simplified four-parameter model, called the truncated power-law model, was introduced by \cite{Lavrov_2015}. This model constitutes  a regularization of the power-law, where the cut-off viscosities for low and high shear rates are introduced. The applicability of the truncated power-law rheology in the hydraulic fracture problems was investigated in the work of \cite{Wrobel_2020} using the example of the PKN geometry \citep{Nordgren,Kusmierczyk_2013}. It was proved that the truncated power-law model is a credible alternative to  the Carreau fluid, providing the results that mimic those obtained for the Carreau variant with accuracy sufficient for any practical application. Thus, when analysing hydraulic fracture problems with the truncated power-law fluid one can extend results to the cases of more advanced four-parameter rheological models. Moreover, while using the truncated power-law model we can introduce explicitly not only the values of cut-off viscosities, but also the magnitudes of limiting shear rates at which the cut-offs occur. In this way one can easily adjust the parameters of fluid rheology and examine precisely their influence on the overall process.

In this paper we analyse a problem of a hydraulic fracture driven by a shear-thinning truncated power-law fluid. The KGD fracture geometry \citep{Wrobel_2015} is considered in a formulation that accounts for the hydraulically induced tangential tractions on the crack flanks \citep{Wrobel_2017,Wrobel_2018,Papanastasiou_2018}. The paper is in a sense complementary to the previous report by \cite{Wrobel_2020}, where it was shown among other things that: i) the truncated power-law is a good alternative for the Carreau model, ii) the fluid flow inside the fracture evolves from the high shear rate Newtonian regime at initial times towards the intermediate shear rate regimes at later stages. As the PKN geometry does not reflect properly the near-tip region of a planar fracture, it is the present research that provides relevant information on the spatial distribution of flow regimes inside the crack, including the near-tip zone. Moreover, in \cite{Wrobel_2020} it was found that the range of shear rates over which the viscosity gradation takes place is equally important for the HF process as the magnitudes of cut-off viscosities themselves. Now, with the truncated power-law we will quantify precisely this influence.

\section{General relations}
\label{gen_rel}
\begin{figure}[htb!]
\begin{center}
\includegraphics[scale=0.55]{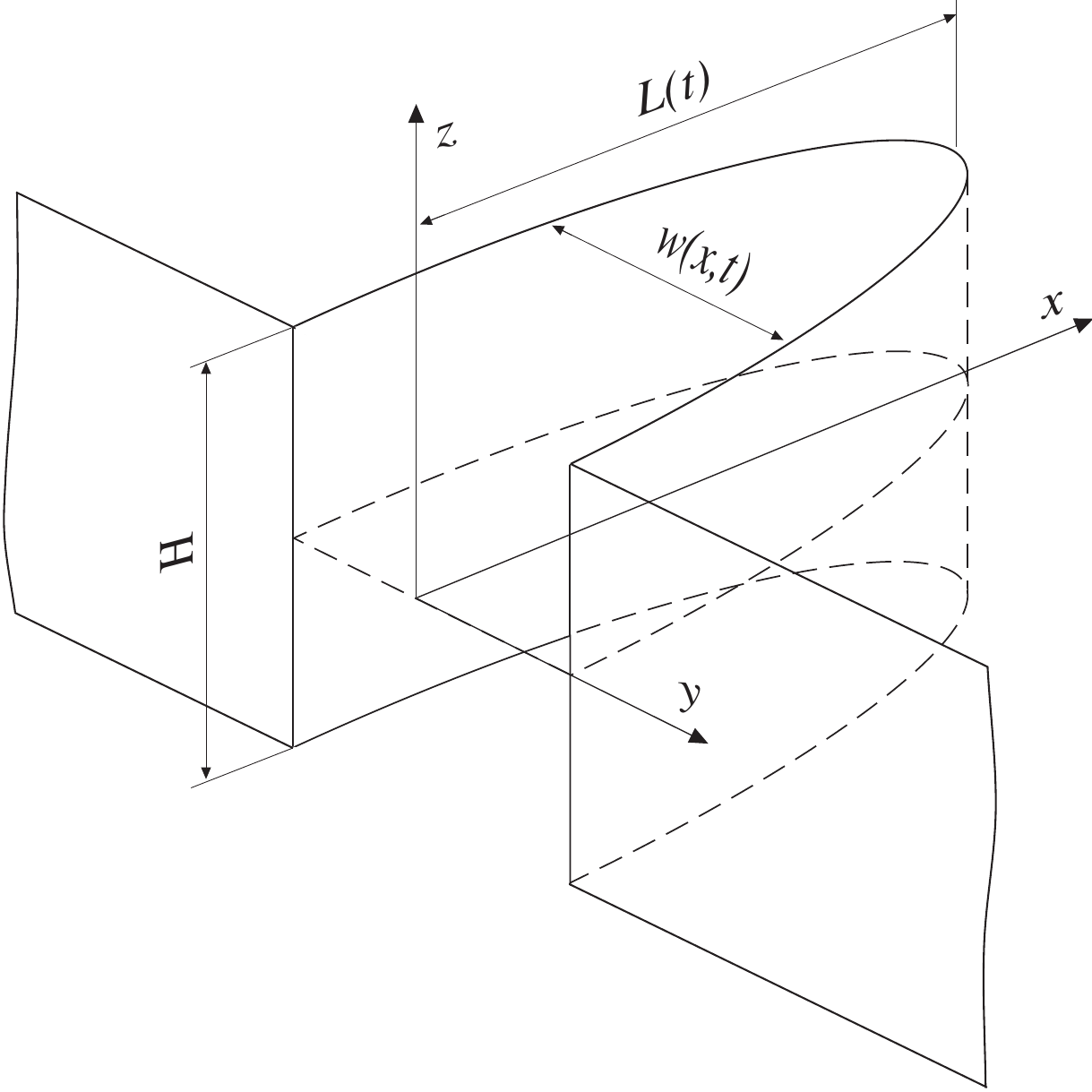}
\caption{The KGD fracture geometry.}
\label{KGD_geom}
\end{center}
\end{figure}

We consider a hydraulic fracture problem based on the formulation introduced by \citet{Wrobel_2017}. It employs the KGD fracture geometry  as schematically shown in Fig. \ref{KGD_geom}. The symmetrical two-winged fracture of length $2L$ propagates in the plane $x \in[-L,L]$, where $L=L(t)$. Our analysis focuses only on one of the symmetrical parts  ($x \in[0,L]$). The fracture height, $H$, is assumed constant, while the fracture opening, $w(x,t)$,  is an element of the solution. 

 The local mass balance within the fracture is described by the continuity equation:
\begin{equation}
\label{cont}
\frac{\partial w}{\partial t}+\frac{\partial q}{\partial x}=0,
\end{equation}
where $q(x,t)$ is the normalised (by factor $H$) fluid flow rate through the fracture cross sections and the solid is assumed to be impermeable (no leak-off). The fluid velocity averaged over the fracture cross section is defined as:
\begin{equation}
\label{v_gen}
v=\frac{q}{w}.
\end{equation}
We assume that there is no lag between the fluid front and the fracture tip and the leak-off is bounded at the crack tip. Thus, the fluid velocity at the fracture tip equals the crack propagation speed:
\begin{equation}
\label{SE}
v(L,t)=\frac{\text{d}L}{\text{d}t}.
\end{equation}

The elastic deformation of the solid material under the applied hydraulic loading is described by the elasticity equation. Here we accept its form introduced in work of \cite{Wrobel_2017}, which accounts for both, the normal pressure, $p$, and the tangential tractions, $\tau$, exerted by the fluid on the crack faces:
\begin{equation}
\label{elasticity_1}
p(x,t)=  \int_{0}^{L(t)} \left[k_2\frac{\partial w(s,t)}{\partial s}-k_1 \tau(s,t) \right] \frac{s\, ds}{x^2 - s^2}, \quad 0\le x<L(t).
\end{equation}
The tangential traction is computed as:
\begin{equation}
\label{tau}
\tau(x,t)=-\frac{w}{2}\frac{\partial p}{\partial x}.
\end{equation}
The respective multipliers $k_1$ and $k_2$ in \eqref{elasticity_1} are:
\begin{equation}
\label{k_1_2}
k_1=\frac{1-2\nu}{\pi(1-\nu)}, \quad k_2=\frac{1}{2\pi}\frac{E}{1-\nu^2},
\end{equation}
with $\nu$ being the Poisson's ratio and $E$ denoting the Young modulus.

In the paper of \cite{Wrobel_2017} it was proved that, when using the elasticity operator \eqref{elasticity_1}, the only permissible type of tip asymptotics is the one of the so-called toughness dominated regime (see also \cite{Wrobel_2015}):
\begin{equation}
\label{w_asymp}
w(x,t)=w_0(t)\sqrt{L(t)-x}+w_1(t)(L(t)-x)+w_2(t)(L(t)-x)^{3/2}\ln(L(t)-x)+...,\quad x\to L(t),
\end{equation}
\begin{equation}
\label{p_asymp}
p(x,t)=p_0(t)\ln(L(t)-x)+p_1(t)+p_2(t)\sqrt{L(t)-x}+p_3(t)(L(t)-x)\ln(L(t)-x)+...,\quad x\to L(t),
\end{equation}
where the respective multipliers $w_i$, $p_i$ depend on time only.

Moreover, when accounting for the hydraulically induced tangential traction, the standard LEFM crack propagation condition no longer holds. Instead, the fracture extension condition based on the Energy Release Rate (ERR) yields \citep{Wrobel_2017}:
\begin{equation}
\label{ERR_cond}
K_I^2+4(1-\nu)K_I K_\text{f}=K_{I\text{c}}^2,
\end{equation}
where $K_I$ is the standard Mode I stress intensity factor, $K_\text{f}$ is the so-called shear-stress intensity factor and $K_{I\text{c}}$ denotes the material fracture toughness. $K_I$ and $K_\text{f}$ are interrelated in the following way:
\begin{equation}
\label{K_f}
K_\text{f}=\frac{p_0}{G-p_0}K_I,
\end{equation}
with $p_0$ being the multiplier of the leading asymptotic term of fluid pressure (compare with \eqref{p_asymp}) and $G$  the bulk shear modulus.

Respective boundary conditions for the problem include:
\begin{itemize}
\item{two tip boundary conditions:
\begin{equation}
\label{BCs_tip}
w(L,t)=0, \quad q(L,t)=0,
\end{equation}}
\item{the influx boundary condition:
\begin{equation}
\label{BC_q}
q(0,t)=q_0(t).
\end{equation}}
\end{itemize}
Finally, the initial conditions define the initial crack length and the initial fracture aperture:
\begin{equation}
\label{init_cond}
L(0)=L_*, \quad w(x,0)=w_*(x).
\end{equation}

\section{Fluid flow equations: truncated power-law model}
\label{fluid_eqs}

In the paper of \cite{Lavrov_2015} the truncated power-law model of fluid was proposed as an alternative for the Carreau law. It constitutes a simple regularization of the pure power-law model, where  low and high shear rate cut-off viscosities, $\eta_0$ and $\eta_\infty$, are employed. In this way, the inherent drawbacks of the power-law rheology are eliminated, whereas a relative ease of computational implementation is retained. In the work of \cite{Wrobel_2020} the results obtained with the truncated power-law fluid were compared with those produced with Carreau model  for the PKN fracture. It turned out that the truncated power-law rheology is indeed a good replacement for the Carreau fluid in the HF problems. It provides the quality of approximation of the Carreau results that is sufficient in any practical application. 

In the truncated power-law model the apparent viscosity is expressed as:
\begin{equation}
\label{TP_def}
\eta_\text{a}=
  \begin{cases}
		\eta_0       & \quad \text{for } \quad |\dot \gamma |<|\dot \gamma_1|,\\
    C |\dot \gamma|^{n-1}       & \quad \text{for} \quad |\dot \gamma_1|<|\dot \gamma|<|\dot \gamma_2|,\\
		
    \eta_\infty  & \quad \text{for } \quad |\dot \gamma|>|\dot \gamma_2|,
  \end{cases}
\end{equation}
where $\dot \gamma$ stands for the fluid shear rate (inside the crack we adopt the coordinate system from  \cite{Wrobel_Arxiv}, as a result $\dot \gamma <0$) , $C$ is the so-called consistency index and $n$ denotes the fluid behaviour index. For $0<n<1$ one obtains  shear-thinning properties, while $n>1$ yields the shear-thickening characteristic. The limiting values of shear rates for which the cut-off viscosities are employed are:
\begin{equation}
\label{gamma_lim}
|\dot \gamma_1|=\left(\frac{C}{\eta_0}\right)^{1/(1-n)}, \quad |\dot \gamma_2|=\left(\frac{C}{\eta_\infty}\right)^{1/(1-n)}.
\end{equation}

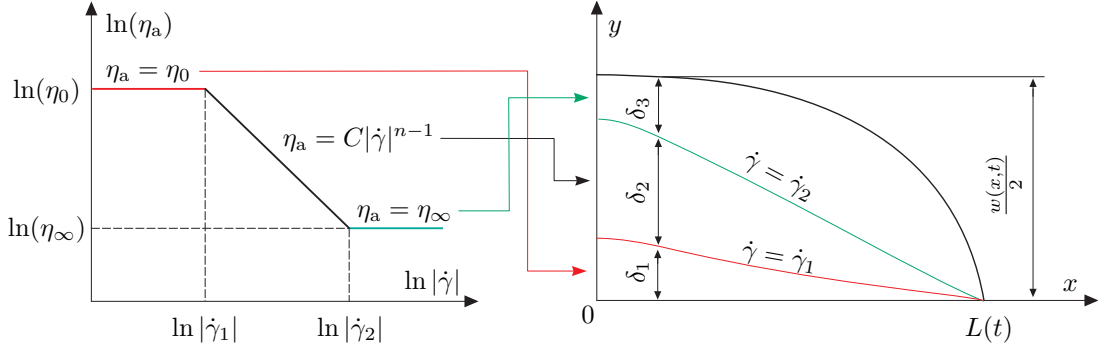
\begin{figure}[htb!]
\begin{center}
\includegraphics[scale=0.7]{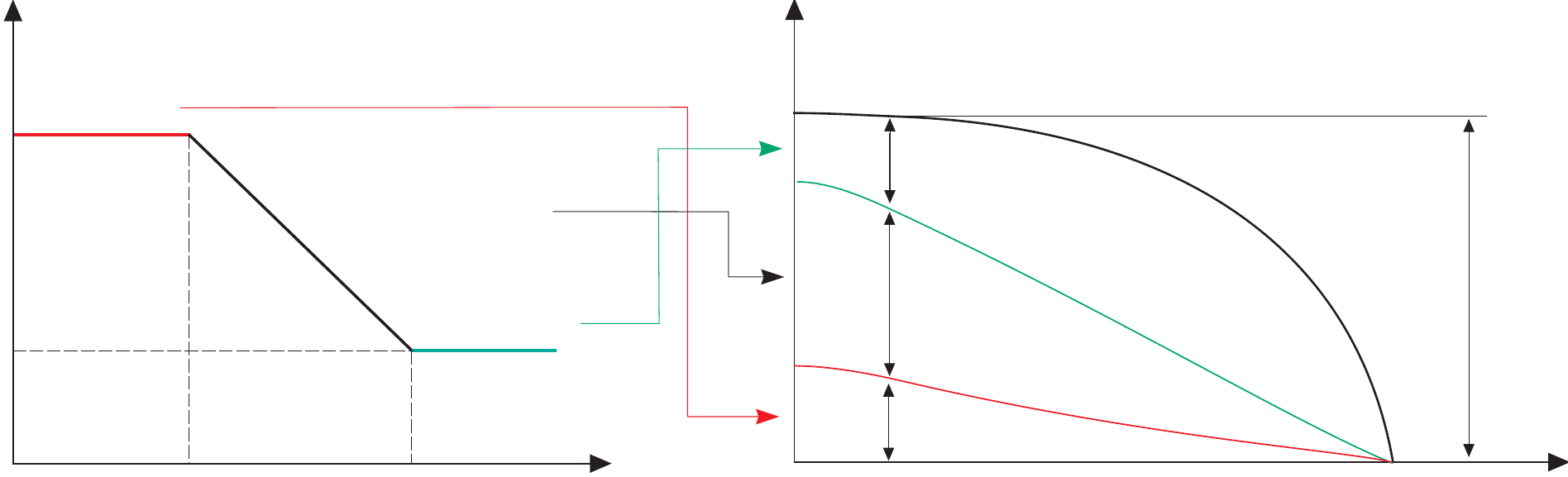}
\put(-13,7){$x$}
\put(-185,105){$y$}
\put(-195,-5){$0$}
\put(-178,10){\rotatebox{90}{$\delta_1$}}
\put(-178,40){\rotatebox{90}{$\delta_2$}}
\put(-178,72){\rotatebox{90}{$\delta_3$}}
\put(-42,35){\rotatebox{90}{$\frac{w(x,t)}{2}$}}
\put(-50,-10){$L(t)$}
\put(-135,20){\rotatebox{-10}{$\dot \gamma=\dot \gamma_1$}}
\put(-135,55){\rotatebox{-28}{$\dot \gamma=\dot \gamma_2$}}
\put(-375,105){$\ln (\eta_\text{a})$}
\put(-262,8){$\ln|\dot \gamma|$}
\put(-375,88){$\eta_\text{a}=\eta_0$}
\put(-280,35){$\eta_\text{a}=\eta_\infty$}
\put(-308,62){$\eta_\text{a}=C|\dot \gamma|^{n-1}$}
\put(-295,-10){$\ln|\dot \gamma_2|$}
\put(-350,-10){$\ln|\dot \gamma_1|$}
\put(-410,80){$\ln (\eta_0)$}
\put(-412,28){$\ln (\eta_\infty)$}
\caption{Interconnection between viscosity characteristics and geometry of respective shear rate layers over the fracture footprint. The thicknesses of shear rate layers change with time and position along the crack length: $\delta_j=\delta_j(x,t)$, ($j=1,2,3$). The limits of shear rate layers (coloured lines) are defined by constant values of $\dot \gamma$, i. e. $\dot \gamma=\dot \gamma_1$ and $\dot \gamma=\dot \gamma_2$, respectively.}
\label{layers_schem}
\end{center}
\end{figure}

When using the truncated power-law rheology to model the flow in a flat channel (slit flow) one obtains up to three shear rate layers in each of the symmetrical parts of the conduit (for a detailed explanation see \cite{Wrobel_Arxiv}):
\begin{itemize}
\item{The low shear rate domain, where the Newtonian-type behaviour of the fluid holds with viscosity $\eta_0$. The layer is  located in the very core of the flow and its thickness is defined by $\delta_1$:
\begin{equation}
\label{delta_1_def}
\delta_1=-\left(\frac{\partial p}{\partial x} \right)^{-1}\eta_0^{\frac{n}{n-1}}C^{\frac{1}{1-n}}.
\end{equation}
If $\delta_1\geq w/2$, then the layer thickness is limited by the fracture width. }
\item{The intermediate shear rate domain with the power-law behaviour of the fluid. Its thickness, $\delta_2$, is:
\begin{equation}
\label{delta_2_def}
\delta_2=-\left(\frac{\partial p}{\partial x} \right)^{-1}C^{\frac{1}{1-n}}\left(\eta_\infty^{\frac{n}{n-1}}- \eta_0^{\frac{n}{n-1}}\right).
\end{equation} 
Again, if $\delta_1+\delta_2\geq w/2$, then the upper boundary of the layer is defined by the crack wall. }
\item{The high shear rate domain that covers this part of the fracture for which  $|\dot \gamma|>|\dot \gamma_2|$. Its thickness is described by $\delta_3$:
\begin{equation}
\label{delta_3_def}
\delta_3=\frac{w}{2}-\delta_1-\delta_2=\frac{w}{2}+\left(\frac{\partial p}{\partial x} \right)^{-1}C^{\frac{1}{1-n}}\eta_\infty^{\frac{n}{n-1}}.
\end{equation}
This layer, if it exists, is adjacent to the crack flank.  The Newtonian model of the fluid with viscosity $\eta_\infty$ is valid here.}
\end{itemize}
The interrelation between the viscosity characteristics $\eta_\text{a}(\dot \gamma)$ and the geometry or respective shear rate layers over the fracture footprint is schematically shown in Fig. \ref{layers_schem}. 

As seen from relations \eqref{delta_1_def}--\eqref{delta_3_def}, for predefined fluid rheology the sizes of respective layers  depend on the values of $w$ and $\partial p/\partial x$. Naturally, they change in space and time as the fracture evolves. Specifically, when approaching the crack tip it is the high shear rate Newtonian layer that tends to occupy the entire width of the fracture with $\partial p/\partial x \to -\infty$ . When estimating the relative (with respect to the crack opening, $w$) thickness of this layer one gets:
\begin{equation}
\label{d_3_est}
\frac{2\delta_3}{w}=1+2C^{\frac{1}{1-n}}\eta_\infty^{\frac{n}{n-1}}\left(w\frac{\partial p}{\partial x}\right)^{-1},
\end{equation}
which combined with the tip asymptotics \eqref{w_asymp}--\eqref{p_asymp} yields:
\begin{equation}
\label{d_3_est_1}
\frac{2\delta_3}{w}=1-2w_0 p_0 C^{\frac{1}{1-n}}\eta_\infty^{\frac{n}{n-1}}\sqrt{L-x}+...,\quad x \to L(t).
\end{equation}

The fluid flow rate for the truncated power-law model can be computed by analytical integration of respective flow equations  \citep{Wrobel_Arxiv}. Depending on the number of existing shear rate layers one has:
\begin{itemize}
\item{for the case of a single (low shear rate) layer:
\begin{equation}
\label{Q_1}
q=-\frac{1}{12\eta_0}\frac{\partial p}{\partial x}w^3,
\end{equation}}
\item{for the case of two (low and intermediate shear rate) layers:
\begin{equation}
\label{Q_2}
q=\frac{2(1-n)}{3(1+2n)}\left( \frac{\partial p}{\partial x}\right)^{-2}C^\frac{3}{1-n}\eta_0^\frac{2n+1}{n-1}+\frac{2n}{2n+1}\left( -\frac{1}{C}\frac{\partial p}{\partial x}\right)^\frac{1}{n}\left(\frac{w}{2} \right)^\frac{2n+1}{n},
\end{equation}}
\item{for the case of three (low, intermediate and high shear rate) layers:
\begin{equation}
\label{Q_3}
q=-\frac{1}{12\eta_\infty}\frac{\partial p}{\partial x}w^3+\frac{2(1-n)}{3(1+2n)}\left(\frac{\partial p}{\partial x}\right)^{-2}C^\frac{3}{1-n}\left(\eta_0^\frac{2n+1}{n-1}-\eta_\infty^\frac{2n+1}{n-1} \right).
\end{equation}}
\end{itemize}

Following \cite{Wrobel_2020} we adopt here a general definition of the fluid flow rate in the form:
\begin{equation}
\label{q_car_def}
q=-\frac{1}{12\eta_\infty}w^3\frac{\partial p}{\partial x}F\left(x,t \right),
\end{equation}
where $F$ can be deduced from \eqref{Q_1}--\eqref{Q_3}. If the shear rates are sufficiently high and consequently the produced apparent viscosity yields $\eta_\text{a}=\eta_\infty$, the function $F$ assumes a unit value. On the other hand, for sufficiently low shear rates (resulting in apparent viscosity $\eta_\text{a}=\eta_0$), $F$ gives $\eta_\infty/\eta_0$.  In this way   $F(x,t)$ provides the information on the extent of deviation of solution from  the high shear rate Newtonian regime of flow in a certain spatial and temporal location. Note that, in the light of estimations \eqref{d_3_est}--\eqref{d_3_est_1}, the asymptotic definition of $F(x,t)$ produces:
\begin{equation}
\label{F_asymp_1}
F(x,t)=1-\frac{12 \eta_\infty A}{\left(w\frac{\partial p}{\partial x}\right)^3}, \quad x \to L(t),
\end{equation}
where:
\[
A=\frac{2(1-n)}{3(1+2n)}C^\frac{3}{1-n}\left(\eta_0^\frac{2n+1}{n-1}-\eta_\infty^\frac{2n+1}{n-1} \right).
\]
By employing the asymptotic relations \eqref{w_asymp}--\eqref{p_asymp} in \eqref{F_asymp_1} we arrive at the following estimation:
\begin{equation}
\label{F_asymp_2}
F(x,t)=1+\frac{12 \eta_\infty A}{w_0^3p_0^3}(L-x)^{3/2}+..., \quad x \to L(t),
\end{equation}
which means that the high shear rate Newtonian regime of flow  is retained in the immediate proximity of the fracture tip. Clearly, the size of the zone over which this asymptotics holds varies with fracture evolution.

\section{Numerical analysis}
\label{num_an}

The rheological properties of fracturing fluid affect crucially the process of hydraulic fracture. Viscosity of many fluids used in fracking treatments can be properly described by four-parameter models such as Carreau and Cross \citep{Lecampion_2018}. Unfortunately, no closed form solutions for fluid flow rate exist for these models, which makes them cumbersome in numerical implementation. In the work of \cite{Wrobel_2020} it was shown that the truncated power-law model can be a good substitute for the Carreau rheology, while simultaneously providing a relative simplicity in numerical implementation. 

In the following analysis we will use the truncated power-law model to investigate the spatial and temporal evolution of the fluid flow regime inside the fracture and the resulting influence on crack geometry. The aim here is not to reproduce rather trivial conclusion that by increasing the fluid viscosity one obtains shorter and wider fracture (and, conversely, when the viscosity is reduced the crack grows longer and thinner). Instead, the core of this study will be concentrated on the investigation how varying values of limiting shear rates affect the process of hydraulic fracturing for fixed values of the cut-off viscosities $\eta_0$  and $\eta_\infty$. The reader interested in the results of variation of $\eta_0$  and $\eta_\infty$ is refereed to the paper by \cite{Wrobel_2020} where the related effects have been quantified and discussed. 

We will consider some hypothetical fluid with limiting viscosities $\eta_0=0.5$ Pa$\cdot$s and $\eta_\infty=10^{-3}$ Pa$\cdot$s. Similar values hold, for example, for the Hydroxypropylguar (HPG)  fracturing fluid \citep{Lecampion_2018}. We set the lower limiting shear rate to $|\dot \gamma_1|=1$ s$^{-1}$, which is also a figure close to that for the HPG fluid. With the above fluid parameters maintained constant we will consider six different magnitudes of $|\dot \gamma_2|$ (see Tab. \ref{T1}). The resulting effects will be quantified and analyzed.

The aforementioned strategy has been adopted for two reasons:
\begin{itemize}
\item{By varying $\dot \gamma_1$ simultaneously with $\dot \gamma_2$ one multiplies the number of variants to be considered during analysis. Thus, the results would be far less legible and more difficult to interpret. On the other hand, respective underlying mechanisms that govern transition between the regimes of flow can be successfully presented even with constant (and properly selected) lower limiting shear rate.}
\item{The assumed value of $|\dot \gamma_1|$ is relatively large when considering properties of the fracturing fluid such as slickwater (compare e.g. \cite{Wrobel_2020}). In this way it produces proportionally large influence of the low shear rate viscosity plateau on the final results. Nevertheless, the computations show that even with such an assumption the low shear rate viscosity plateau affects the results in a rather minor way. Thus, when decreasing  $|\dot \gamma_1|$ in sensitivity analysis one would see virtually no difference in results with respect to those presented in this paper. On the other hand, increasing $|\dot \gamma_1|$ does not seem justified as the value already accepted is rather large.}
\end{itemize}

\begin{figure}[htb!]
\begin{center}
\includegraphics[scale=0.55]{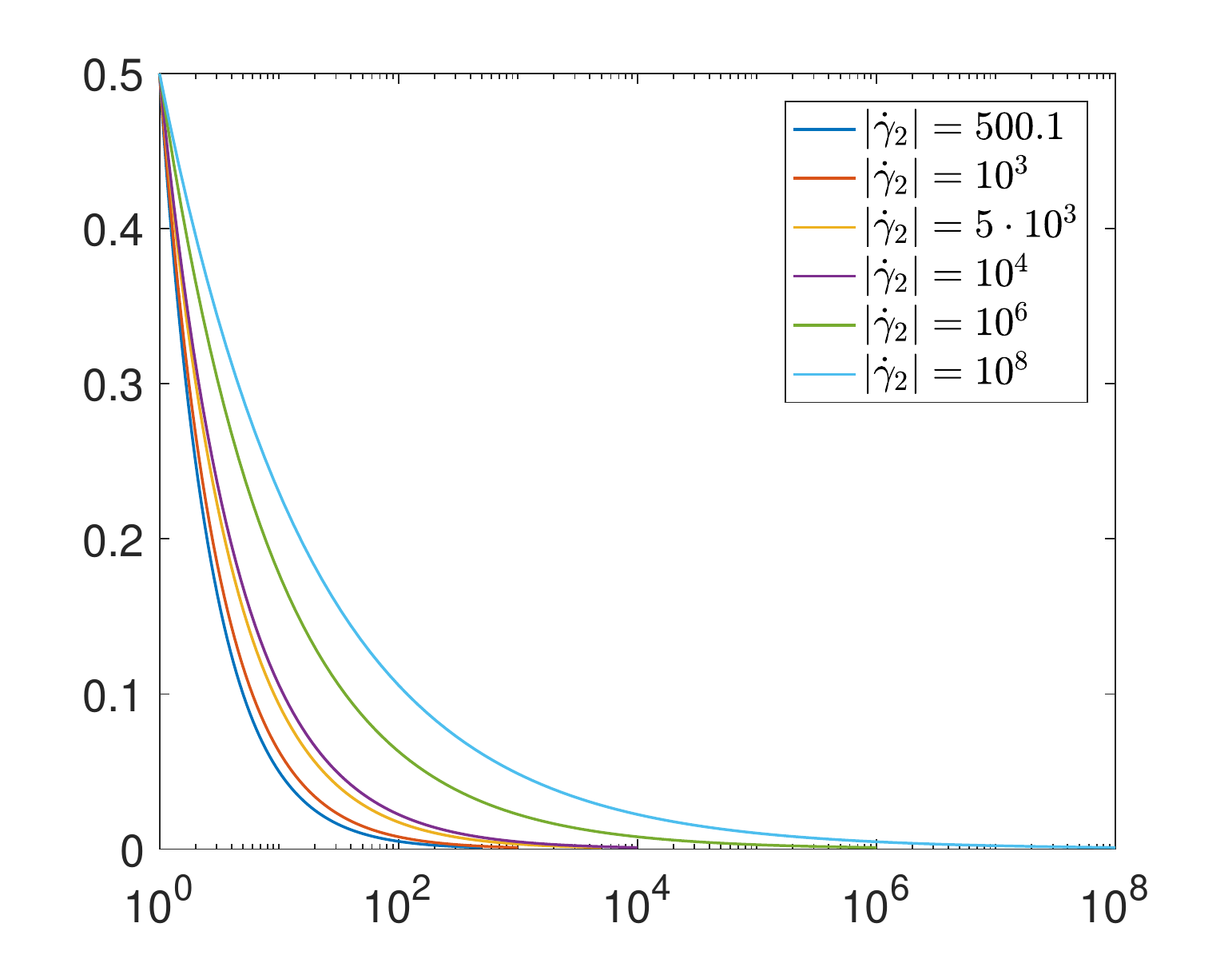}
\put(-125,-5){$|\dot \gamma|$}
\put(-250,90){$\eta_\text{a}$}
\caption{The power-law section of the viscosity characteristics for different values of $|\dot \gamma_2|$.}
\label{eta_gam}
\end{center}
\end{figure}

 Note that for each considered variant the parameters of the truncated power-law model \eqref{TP_def} are computed as:
\begin{equation}
\label{C_n_comp}
n=\frac{\log(\eta_0/\eta_1)}{\log(\dot \gamma_1/\dot \gamma_2)}+1, \quad C=\frac{\eta_0}{|\dot \gamma_1|^{n-1}}=\frac{\eta_\infty}{|\dot \gamma_2|^{n-1}}.
\end{equation}
In order to obtain shear-thinning behaviour of the fluid ($0<n<1$) the following condition has to be satisfied:
\begin{equation}
\label{n_cond}
|\dot \gamma_2|>\frac{\eta_0}{\eta_\infty}|\dot \gamma_1|.
\end{equation}
The lowest value of $|\dot \gamma_2|$ ($500.1$ s$^{-1}$) was taken very close to the limit \eqref{n_cond}\footnote{We do  not want to analyze here the special case of a perfectly plastic fluid, $n=0$, which produces a plug flow. However, the applied methodology can be employed for $n=0$ as well.}. As a result the power-law part of the viscosity characteristics resembles that of a perfectly plastic fluid. On the other hand the highest value of  $|\dot \gamma_2|$ was set two orders of magnitude greater than the one of the HPG fluid. The power law sections of the viscosity characteristics \eqref{TP_def} in respective variants are depicted in Fig. \ref{eta_gam}.

\begin{table}[]
\begin{center}
\begin{tabular}{|c||c|c|c|c|c|c|}
\hline
$\eta_0$, Pa$\cdot$s         & \multicolumn{6}{c|}{$0.5$}                                  \\ \hline
$\eta_\infty$, Pa$\cdot$s    & \multicolumn{6}{c|}{$10^{-3}$}                              \\ \hline
$|\dot \gamma_1|$, s$^{-1}$  & \multicolumn{6}{c|}{$1$}                                    \\ \hline
$|\dot \gamma_2|$, s$^{-1}$  & $500.1$ & $10^3$ & $5\cdot 10^3$ & $10^4$ & $10^6$ & $10^8$ \\ \hline
$C$, Pa$\cdot$s$^{n}$ & \multicolumn{6}{c|}{$0.5$}    \\ \hline
$n$                          & $3.27\cdot10^{-5}$     & $0.1003$    & $0.2703$           & $0.3253$   & $0.5502$   & $0.6626$   \\ \hline
\end{tabular}
\caption{Fracturing fluid parameters.}
\label{T1}
\end{center}
\end{table}

For the above specified cases of the fracturing fluid rheology we will perform a number of simulations assuming some typical values of the HF parameters. Following \cite{Papanastasiou_1999} we set:  $E=16.2$ GPa, $\nu=0.3$, $K_{I\text{c}}=1$ MPa$\cdot\sqrt{\text{m}}$, $q_0=5 \cdot 10^{-4}$ $ \frac{\text{m}^2}{\text{s}} $. The influx magnitude is increased from zero for $t=0$ s to the maximum  $q_0$ at $t_1=1$ s and then kept constant:
\begin{equation}
\label{q0_def}
\bar q_0(t)=
  \begin{cases}
		\left(\frac{3}{t_1^2}t^2-\frac{2}{t_1^3}t^3\right)q_0      & \quad t<t_1,\\
    q_0       & \quad \text{for} \quad t \geq t_1.
  \end{cases}
\end{equation}
Note that the above formula provides a smooth transition between the limiting values of influx. The initial fracture length and velocity are taken zero. The overall time of the process is $t_\text{end}=20$ s. In this time a relatively short fracture is created which agrees with a common assumption that the KGD geometry can be used to describe short cracks \citep{Wrobel_2015}. By extending the time of simulations one would produce the results that exceed the scope of practical applicability of the KGD model. On the other hand,  among the simplified 1D HF models it is the PKN variant which describes better the long fractures. Pertinent analysis for the PKN model has already been done by \cite{Wrobel_2020} where the overall simulation time was set to one hour.

The computations are performed by the HF solver initially developed in \cite{Wrobel_2015,Perkowska_2016}. The subroutine to compute the fluid flow rate for the truncated power-law model was adopted from \cite{Wrobel_Arxiv}.

\begin{figure}[htb!]
\begin{center}
\includegraphics[scale=0.5]{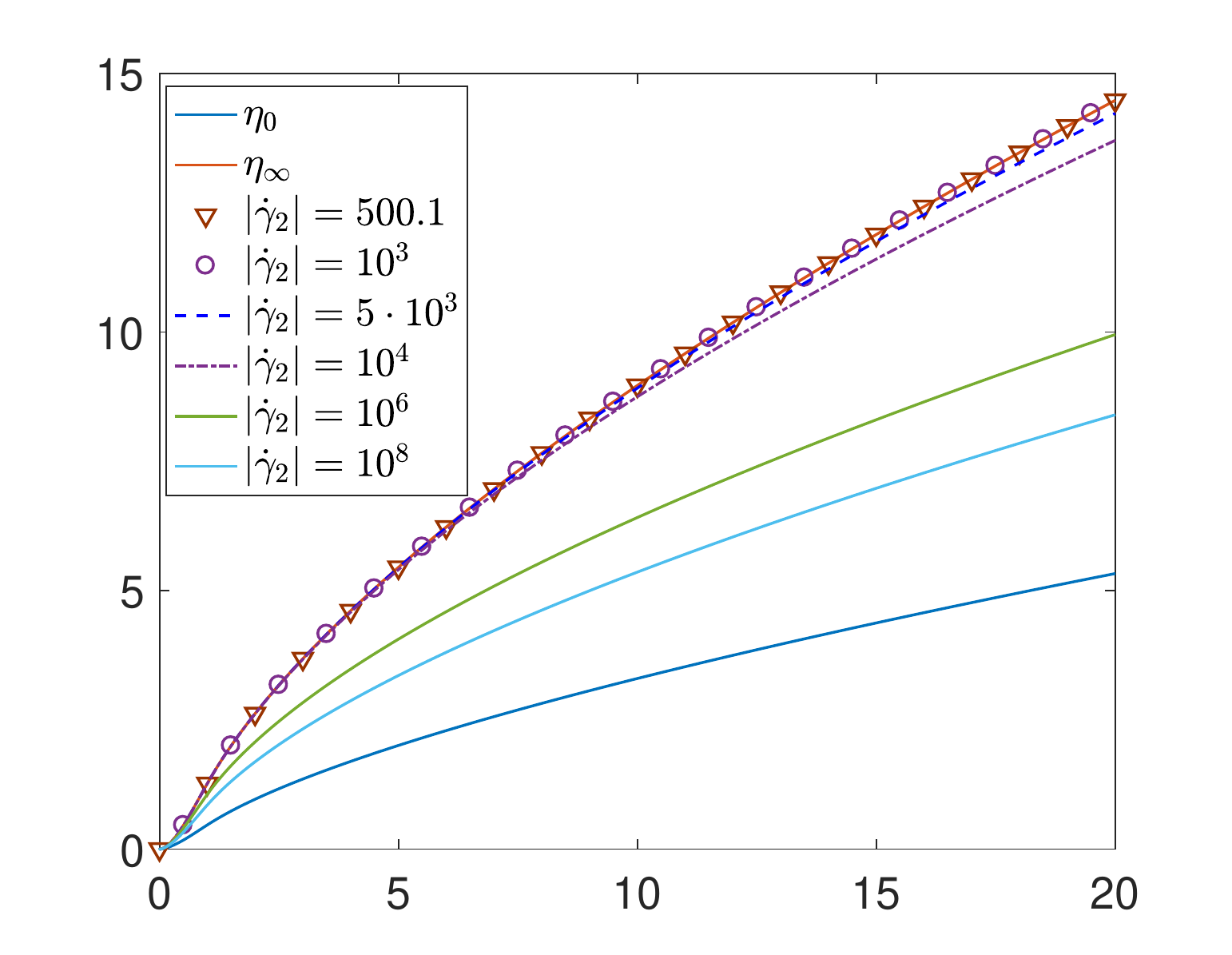}
\includegraphics[scale=0.5]{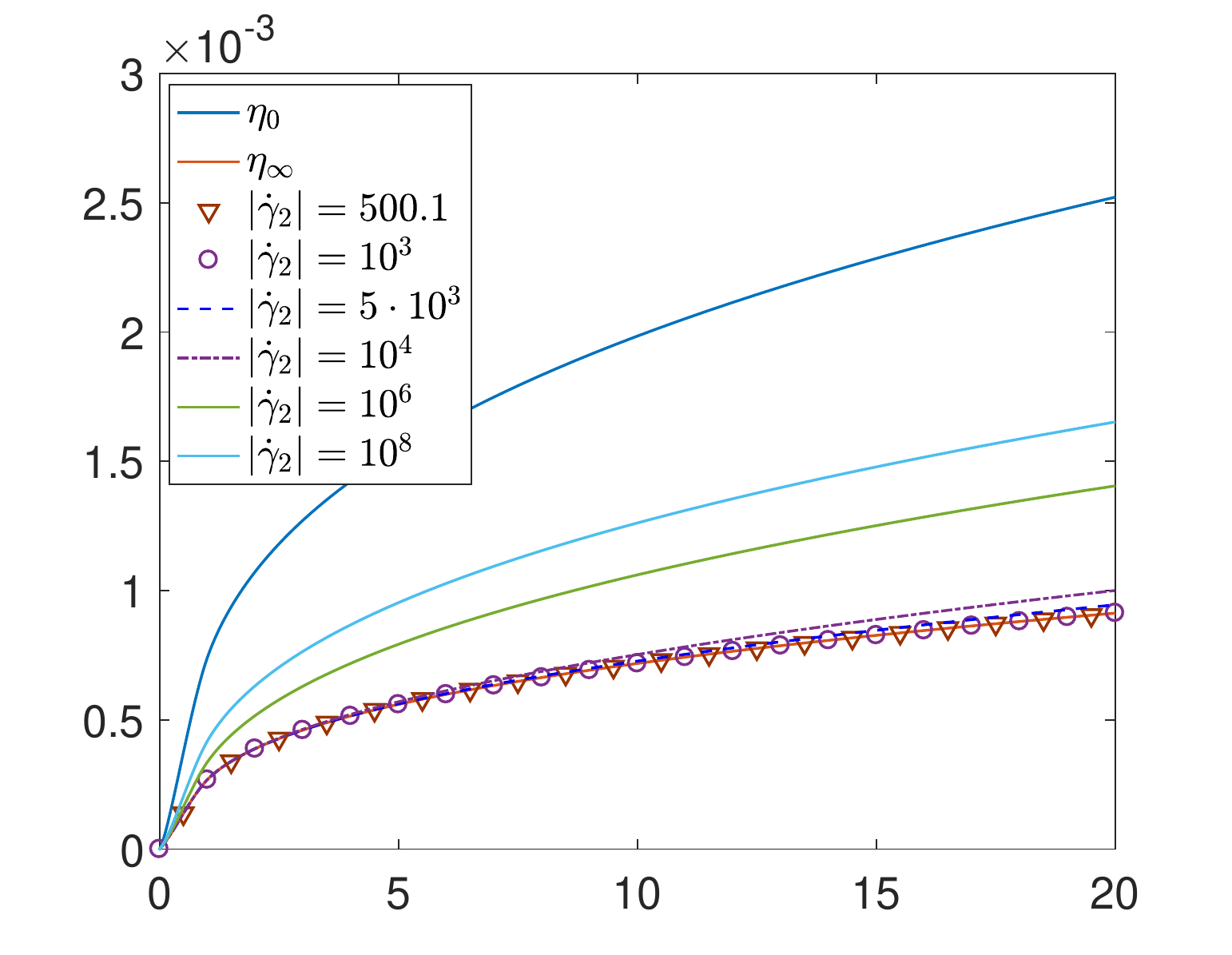}
\put(-338,-5){$t$}
\put(-110,-5){$t$}
\put(-446,165){$\textbf{a)}$}
\put(-220,165){$\textbf{b)}$}
\put(-440,90){$L$}
\put(-240,90){$w(0,t)$}
\caption{Simulation results in terms of: a) the crack length, $L$ [m], and b) the crack opening, $w(0,t)$ [m], at the fracture mouth. Curves denoted as $\eta_0$ and $\eta_\infty$ refer to solutions obtained for the Newtonian fluids of corresponding viscosities. }
\label{L_v0}
\end{center}
\end{figure}

\begin{figure}[htb!]
\begin{center}
\includegraphics[scale=0.5]{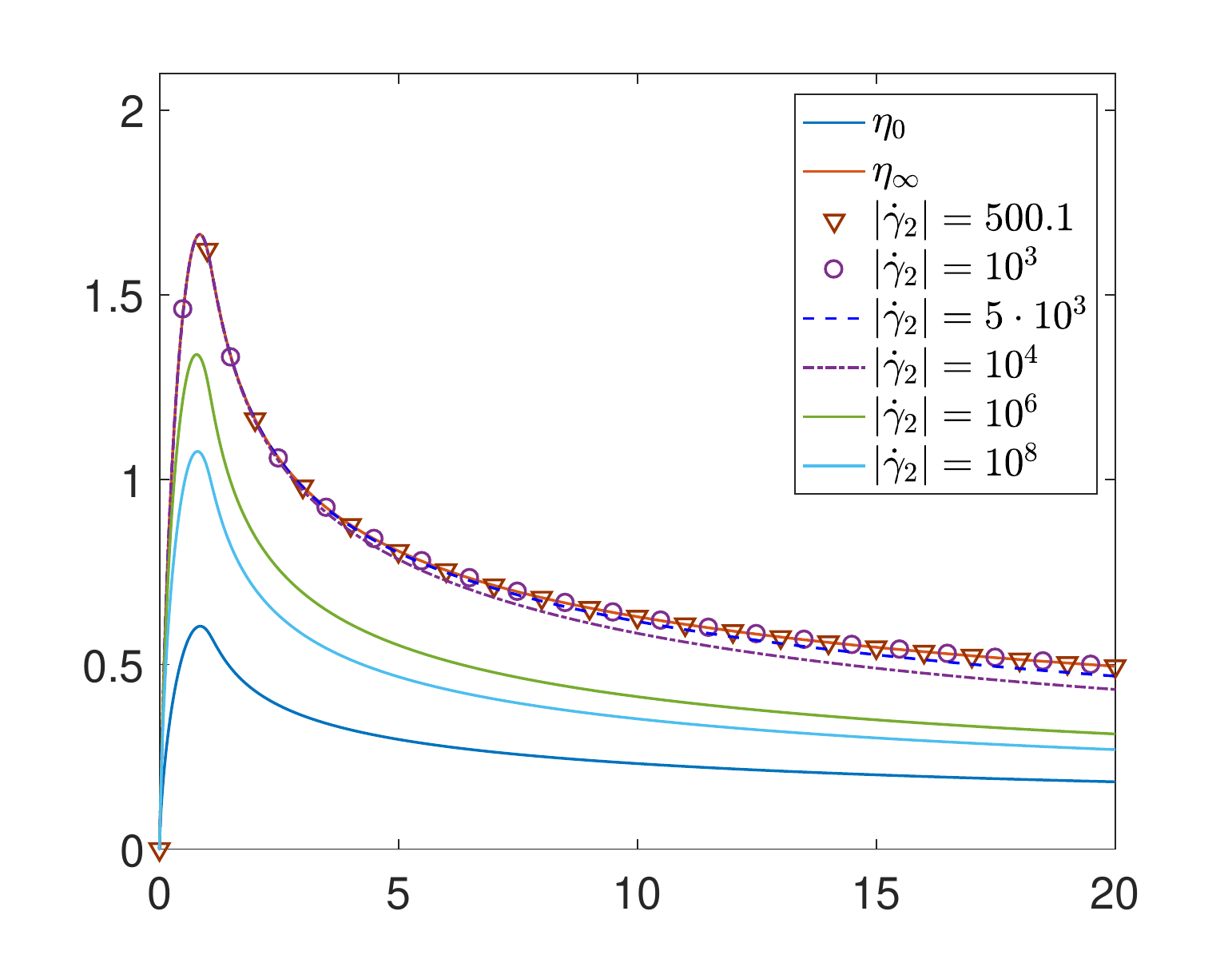}
\includegraphics[scale=0.5]{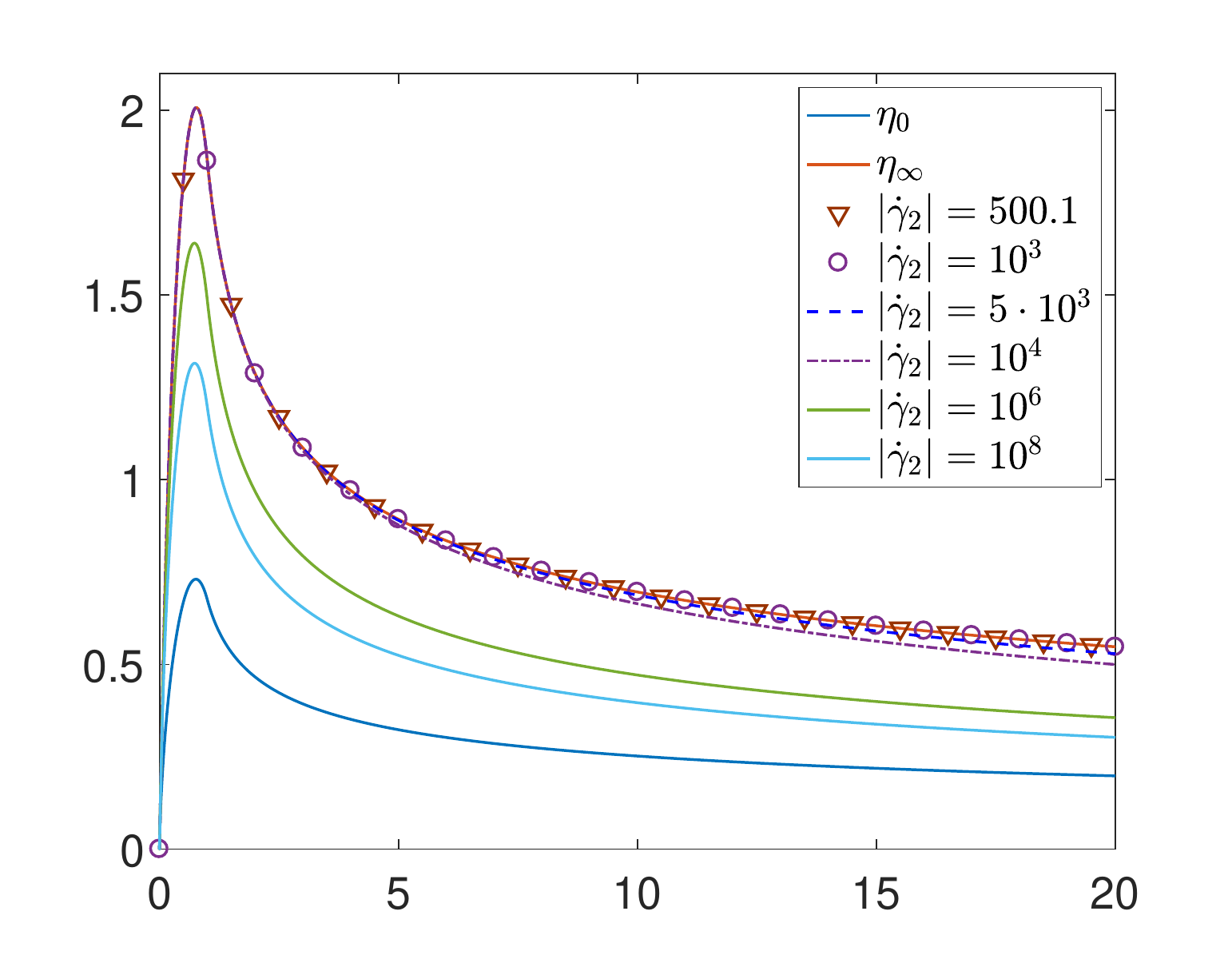}
\put(-338,-5){$t$}
\put(-110,-5){$t$}
\put(-446,165){$\textbf{a)}$}
\put(-220,165){$\textbf{b)}$}
\put(-440,90){$v_0$}
\put(-230,90){$v(0,t)$}
\caption{Simulation results in terms of: a) the crack propagation speed, $v_0$ $\left[\frac{\text{m}}{\text{s}}\right]$, and b) fluid velocity at the fracture mouth, $v(0,t)$ $\left[\frac{\text{m}}{\text{s}}\right]$. Curves denoted as $\eta_0$ and $\eta_\infty$ refer to solutions obtained for the Newtonian fluids of corresponding viscosities. }
\label{wz_vz}
\end{center}
\end{figure}

\begin{figure}[htb!]
\begin{center}
\includegraphics[scale=0.55]{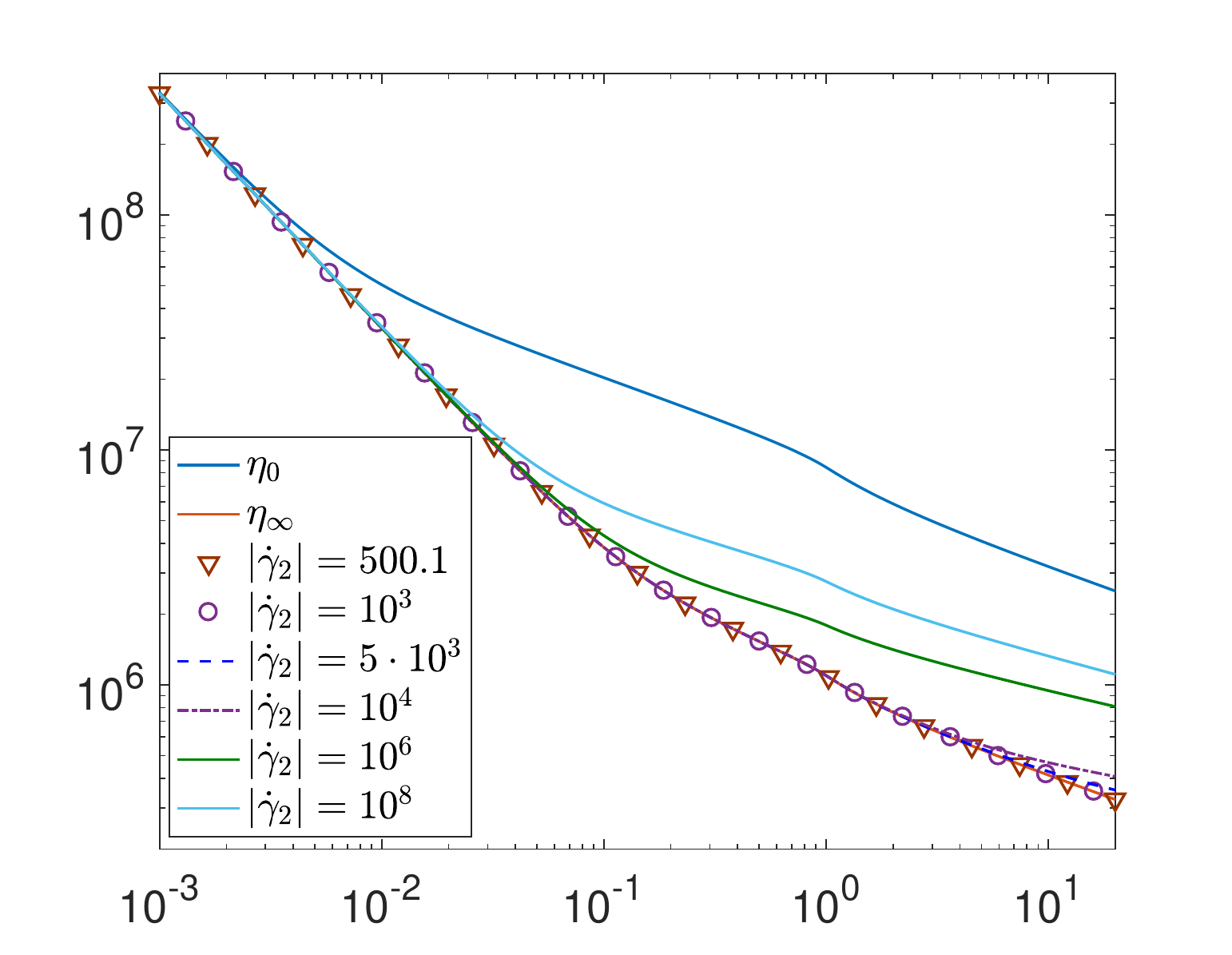}
\put(-125,-5){$t$}
\put(-270,90){$p(0,t)$}
\caption{The downhole net fluid pressure, $p(0,t)$ [Pa]. Curves denoted as $\eta_0$ and $\eta_\infty$ refer to solutions obtained for the Newtonian fluids of corresponding viscosities.}
\label{net_pressure}
\end{center}
\end{figure}

Let us start the analysis of  computational results with some insight into the geometry of created fractures and the fluid flow speed. In Figs. \ref{L_v0}--\ref{net_pressure} we depict the obtained: i) crack lengths, $L$, ii) crack openings at $x=0$, $w(0,t)$, iii) crack propagation speeds, $v_0$,  iv) fluid velocities at the crack mouth, $v(0,t)$, and v) the downhole net fluid pressures, $p(0,t)$. Additionally, we present two solutions produced for Newtonian fluids with viscosites $\eta_0$ and $\eta_\infty$, respectively. Any solution for a shear thinning fluid ($0<n<1$) has to be encompassed by these two limiting  variants. In particular, when $|\dot \gamma_2|$ is sufficiently small, then the results tend to those obtained for $\eta_\infty$ (compare \cite{Wrobel_2020}). On the other hand, with  $|\dot \gamma_2| \to \infty$ we are moving towards the Newtonian solution with viscosity $\eta_0$.

As can be seen in the figures, with growing $|\dot \gamma_2|$ the fracture becomes shorter and wider. Simultaneously, the crack propagation speed and the fluid velocity decrease while the downhole net fluid pressure grows. This rather trivial observation is explained by the fact that growing $|\dot \gamma_2|$ makes the fracturing fluid more viscous over the increasingly longer section of the fracture. Respective data for $|\dot \gamma_2|\leq10^3$ s$^{-1}$ is virtually indistinguishable from the results obtained for $\eta_\text{a}=\eta_\infty$ (and thus we use only markers to depict the former in the figures). This suggests that the fluid flow in these cases is close to the Newtonian high shear rate regime. At the same time, even the solution obtained for $|\dot \gamma_2|=10^8$ s$^{-1}$ is still far away from the low shear rate limiting case ($\eta_\text{a}=\eta_0$). In order to trace the spatial and temporal evolution of the related flow regimes we will investigate below three of $|\dot \gamma_2|$ variants ($|\dot \gamma_2|=\{500.1,5\cdot 10^3,10^6\}$ s$^{-1}$) in some more detail. By using these examples we will show the underlying mechanisms that determine the interrelation between rheological properties of the fluid and the regime of flow.

\subsection{$|\dot \gamma_2|=500.1$ $\text{s}^{-1}$}

\begin{figure}[htb!]
\begin{center}
\includegraphics[scale=0.5]{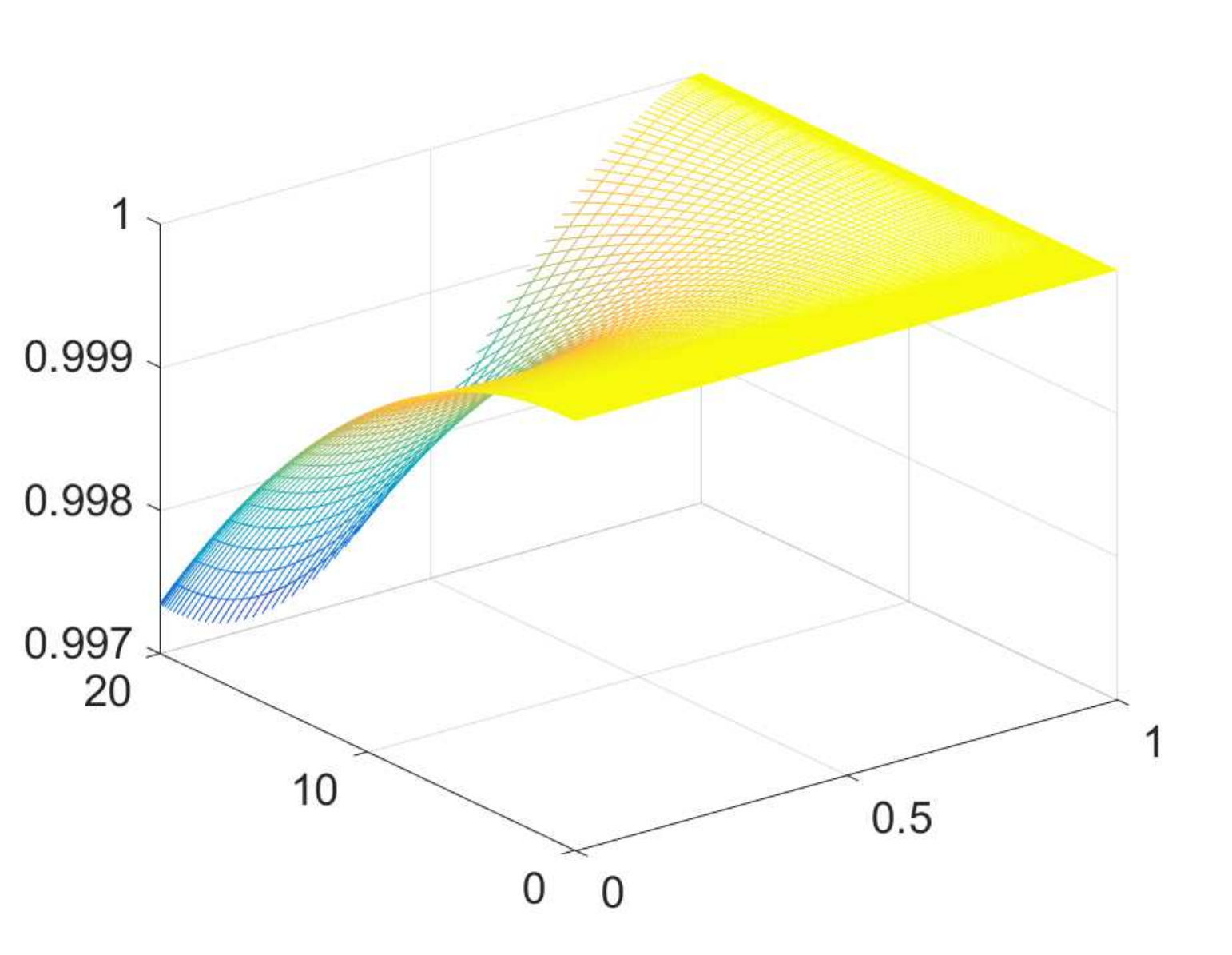}
\put(-60,10){$\frac{x}{L}$}
\put(-180,10){$t$}
\put(-230,80){$F$}
\caption{The component function of fluid flow rate, $F(x,t)$, for $|\dot \gamma_2|=500.1$ s$^{-1}$.}
\label{F_500}
\end{center}
\end{figure}

Let us start by analyzing the fluid flux component function $F(x,t)$ of \eqref{q_car_def}. Its distribution over space and time is depicted in Fig. \ref{F_500}. It shows that $F$ yields 1 virtually over the entire space and time domain. This indicates that the fracture is in the high shear rate Newtonian regime of flow during the whole process of its propagation. For this reason the solution presented in Figs. \ref{L_v0}--\ref{wz_vz} coincides perfectly with that produced for the Newtonian high shear rate model ($\eta_\text{a}=\eta_\infty$).

\begin{figure}[htb!]
\begin{center}
\includegraphics[scale=0.5]{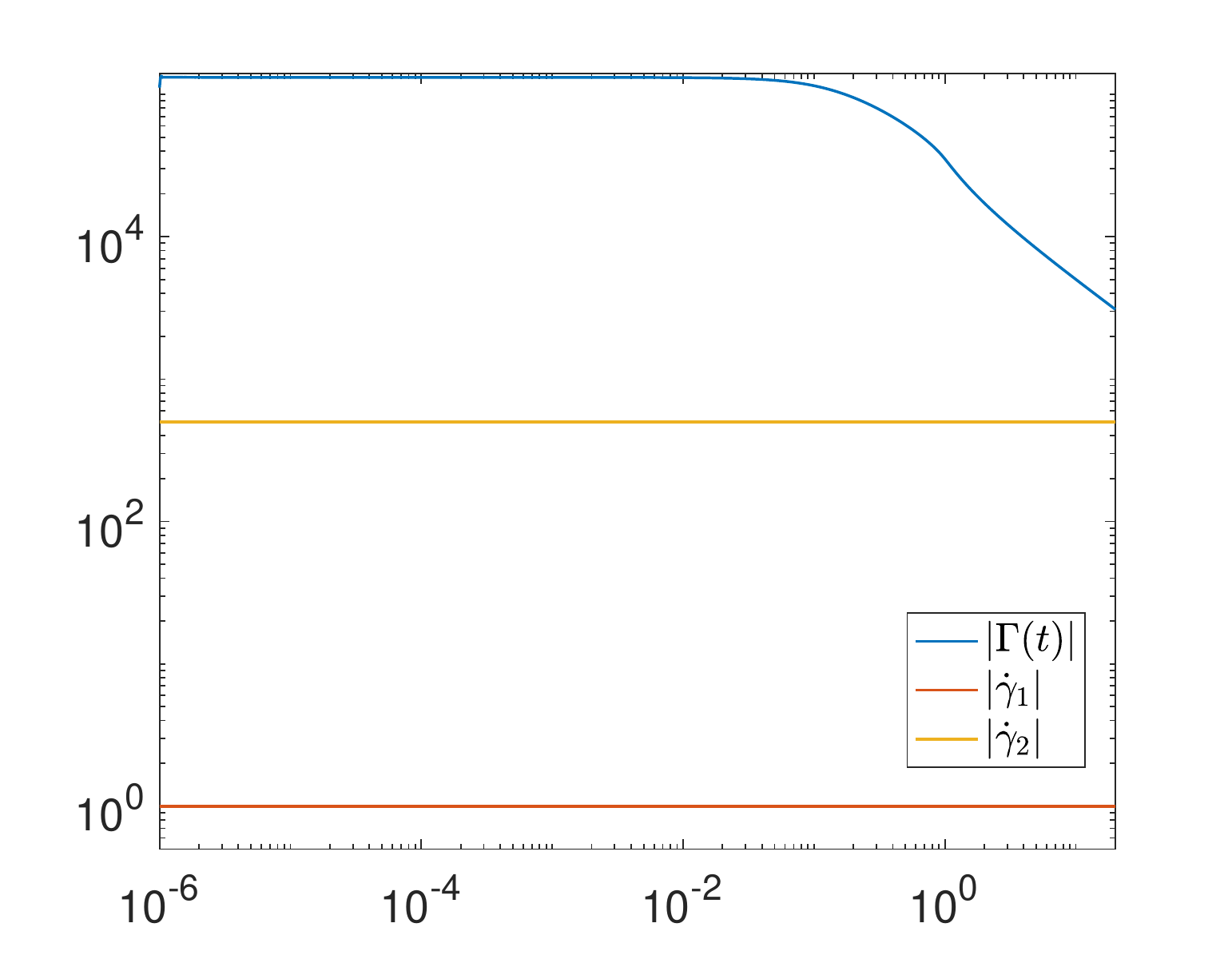}
\put(-120,-5){$t$}
\put(-230,80){$|\Gamma|$}
\caption{The fluid shear rate averaged over fracture surface, $|\Gamma(t)|$, for $|\dot \gamma_2|=500.1$ s$^{-1}$.}
\label{Gam_500}
\end{center}
\end{figure}

In \cite{Wrobel_2020} it was shown that the fluid shear rate integrated over the fracture surface can be a very informative  parameter in understanding the regime of flow. We introduce here the following definition of average shear rate:
\begin{equation}
\label{Gam_def}
\Gamma(t)=\frac{2}{L(t)}\int_0^{L(t)}\frac{1}{w(x,t)}\int_0^{\frac{w(x,t)}{2}}\dot \gamma(x,y,t)\text{d}y\text{d}x.
\end{equation}
The evolution of $|\Gamma(t)|$ in time is presented in Fig. \ref{Gam_500}. For comparison we plotted in the figure also the values of $|\dot \gamma_1|$ and $|\dot \gamma_2|$. One can see that the average shear rates are above the higher cut-off threshold $|\dot \gamma_2|$ over the entire time interval. Thus, even though locally $|\dot \gamma|$ assumes smaller values, it is the high shear rate part of viscosity characteristics that dominates the process. It is also notable that $|\Gamma(t)|$ decreases with time.

\begin{figure}[htb!]
\begin{center}
\includegraphics[scale=0.5]{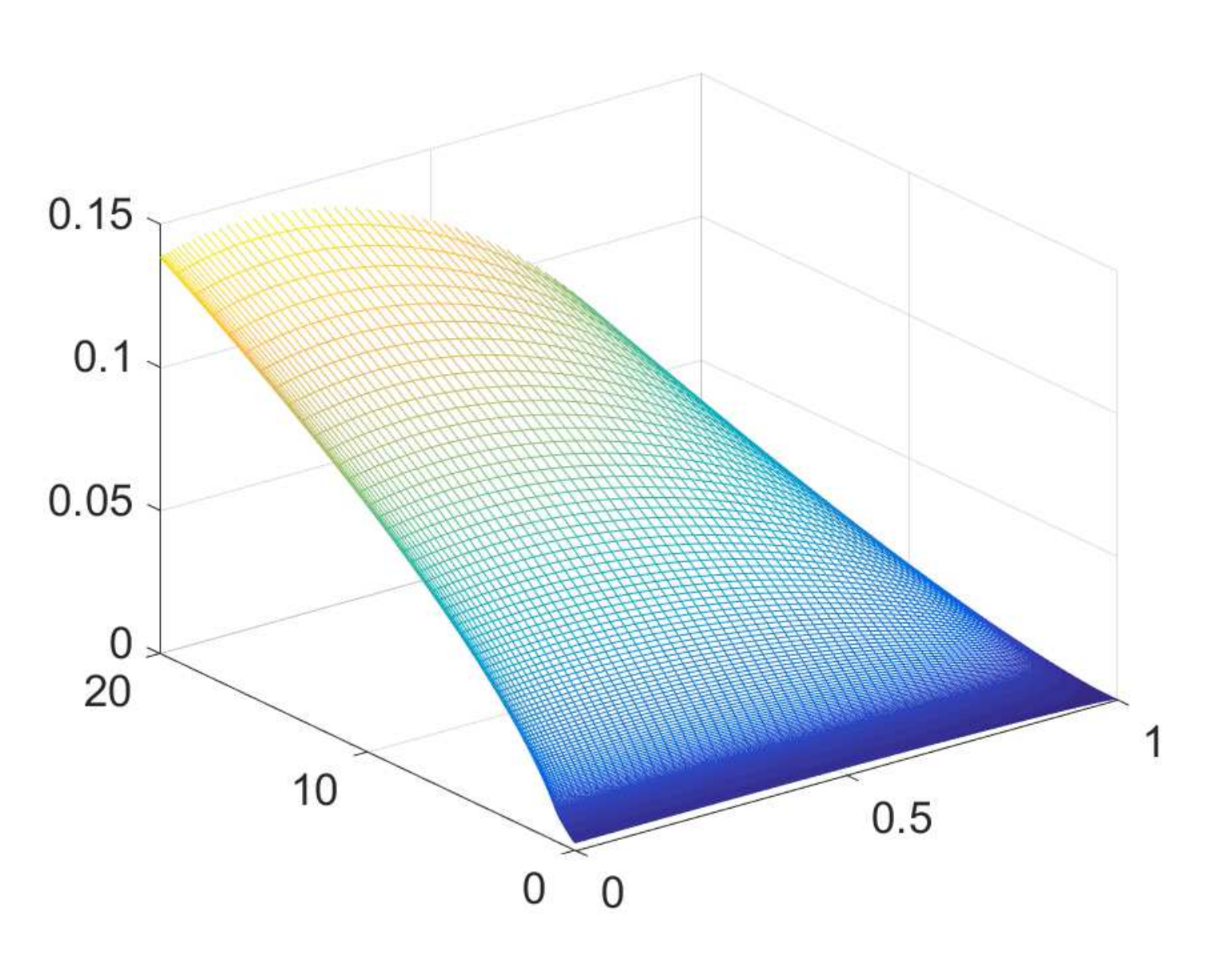}
\put(-60,10){$\frac{x}{L}$}
\put(-180,15){$t$}
\put(-230,95){$\frac{2\delta_1}{w}$}
\includegraphics[scale=0.5]{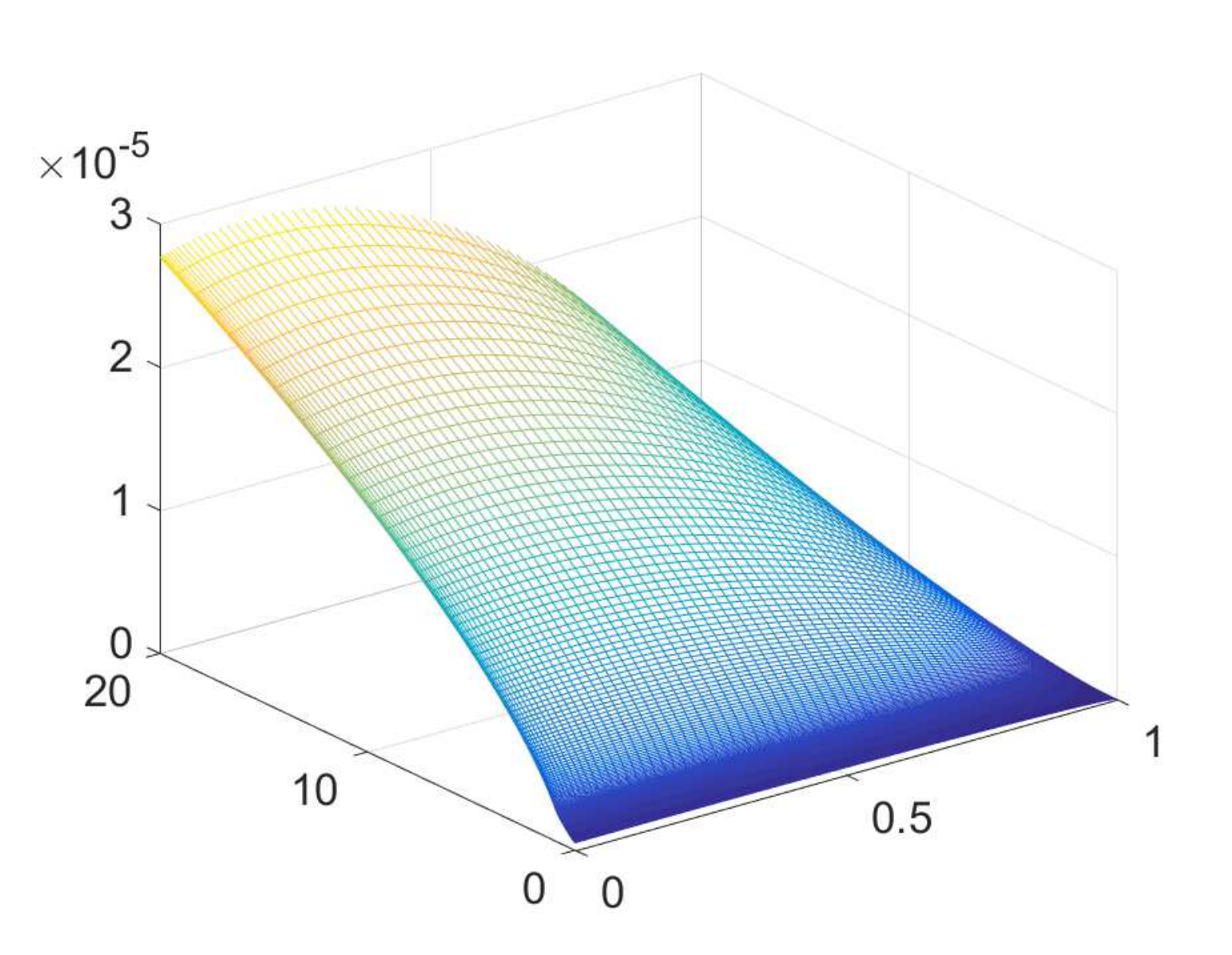}
\put(-60,10){$\frac{x}{L}$}
\put(-180,15){$t$}
\put(-220,95){$\frac{2\delta_2}{w}$}
\put(-446,165){$\textbf{a)}$}
\put(-220,165){$\textbf{b)}$}
\caption{The relative thickness of: a) the low shear rate layer, $\delta_1$, and b) the intermediate shear rate layer, $\delta_2$, for $|\dot \gamma_2|=500.1$ s$^{-1}$.}
\label{d1_d2_500}
\end{center}
\end{figure}
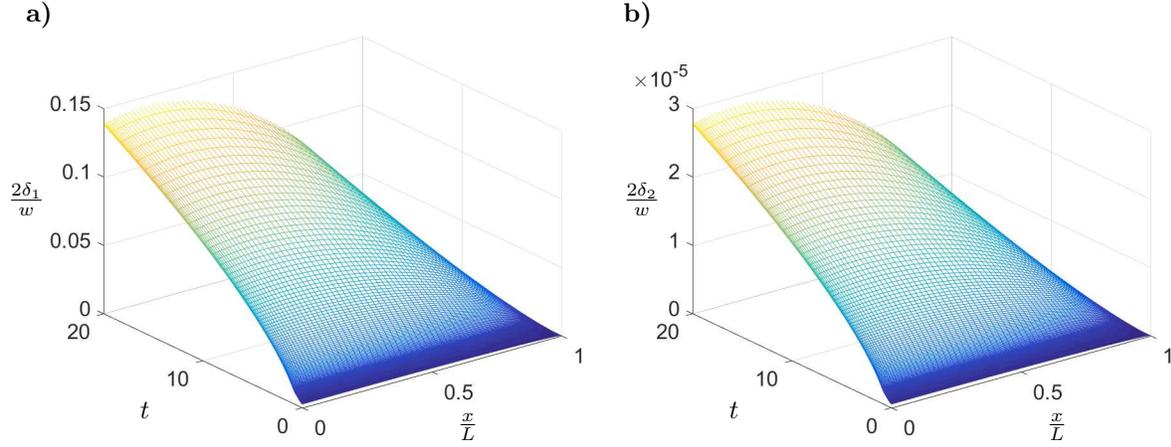

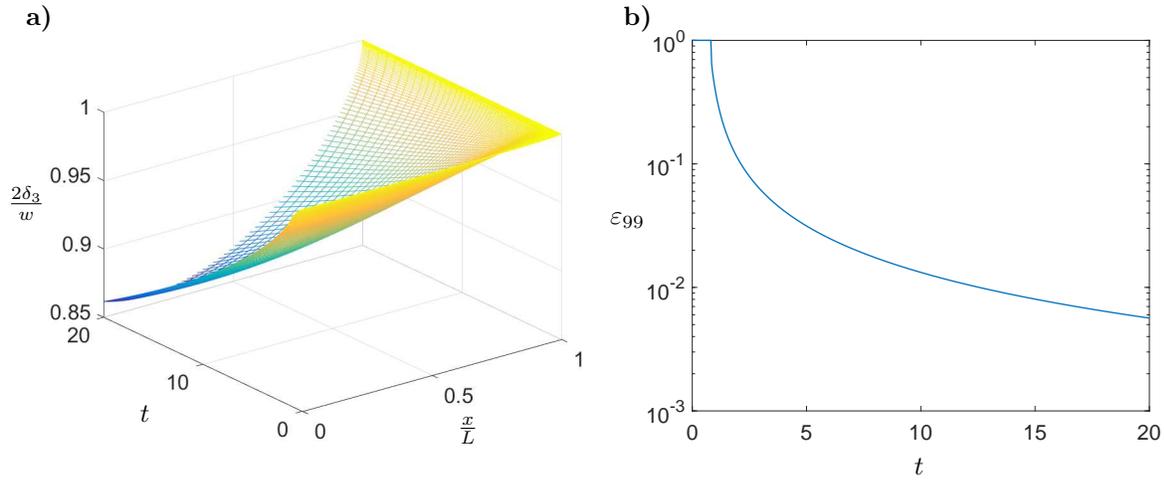
\begin{figure}[htb!]
\begin{center}
\includegraphics[scale=0.5]{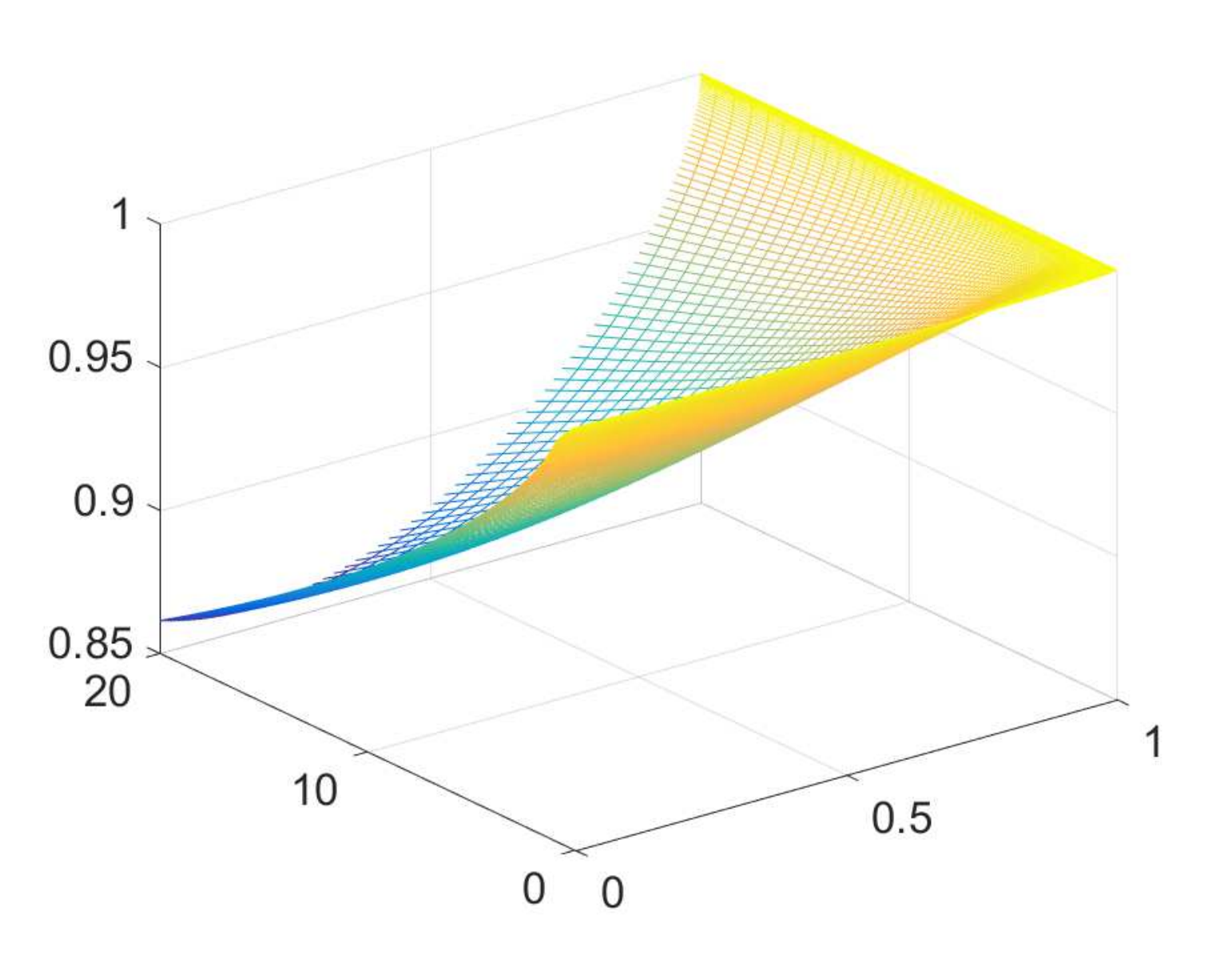}
\put(-60,10){$\frac{x}{L}$}
\put(-180,15){$t$}
\put(-230,95){$\frac{2\delta_3}{w}$}
\includegraphics[scale=0.5]{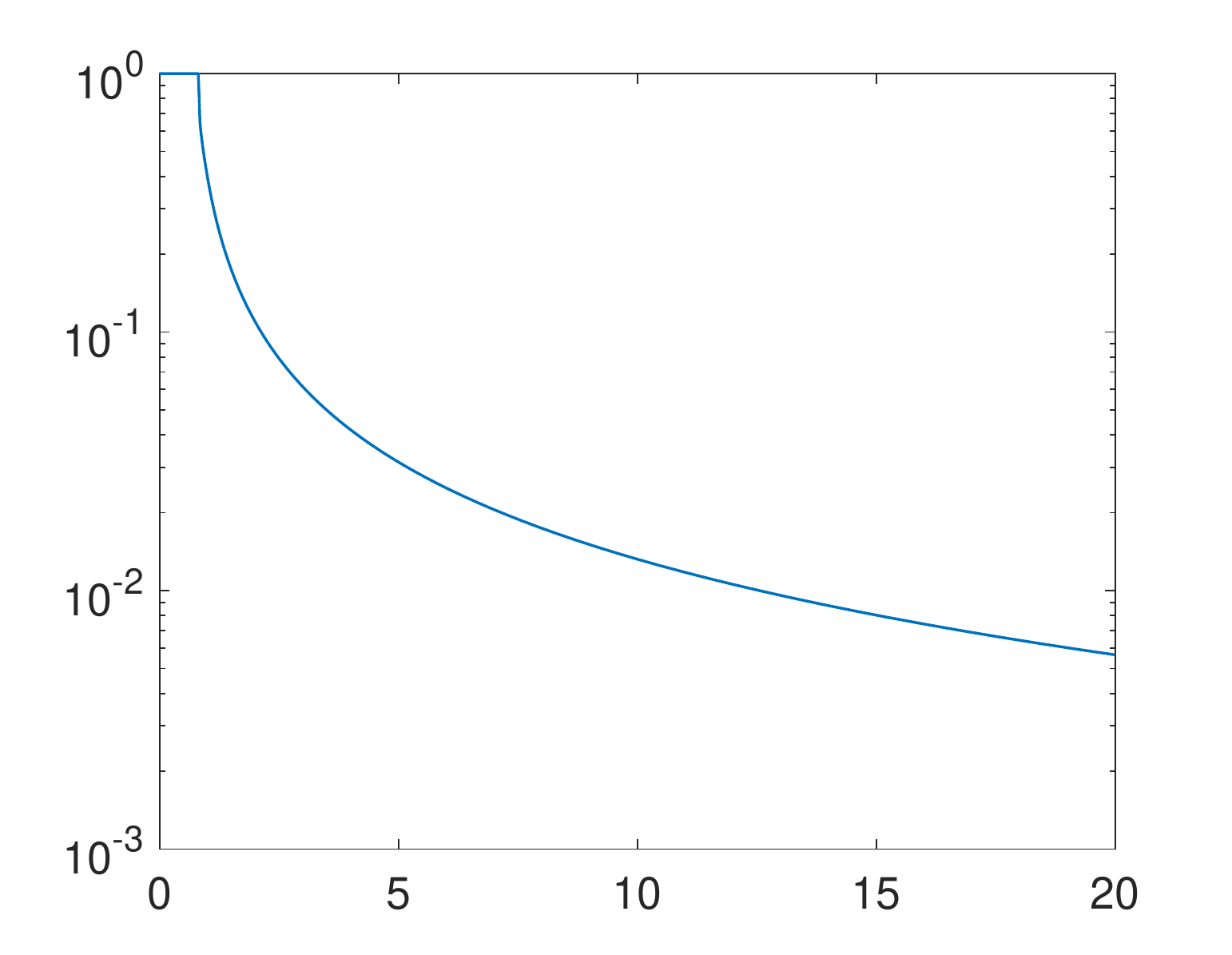}
\put(-110,-5){$t$}
\put(-446,165){$\textbf{a)}$}
\put(-220,165){$\textbf{b)}$}
\put(-225,90){$\varepsilon_{99}$}
\caption{a) The relative thickness of  the high shear rate layer. b) The relative length of the high shear rate Newtonian layer, $\varepsilon_{99}$, for $|\dot \gamma_2|=500.1$ s$^{-1}$.}
\label{d3_d99_500}
\end{center}
\end{figure}

In Fig. \ref{d1_d2_500} and Fig. \ref{d3_d99_500}a) we depict the relative, with respect to the fracture width, thicknesses of corresponding shear rate layers (see \eqref{delta_1_def}--\eqref{delta_3_def}). It turns out that the Newtonian high-shear rate layer occupies over 85 $\%$ of the fracture width at any point of space and time. Obviously, according to the estimation \eqref{F_asymp_2}, this number grows to 100 $\%$ when approaching the crack tip. The Newtonian low shear rate layer ($\eta_\text{a}=\eta_\infty$) only in the final time instant stretches for little less than 15 $\%$ of the crack width in the proximity of the fracture inlet, and then shrinks towards the tip. The size of the intermediate shear rate power-law layer is negligible as compared to the previous two. A clear  trend of $\delta_1$ and $\delta_2$ growth in time at the expense of $\delta_3$ can be noticed.

In order to quantify the size of the near tip high shear rate Newtonian layer we introduce here the characteristic distance from the crack tip, $\varepsilon_{99}$, over which this layer occupies at least 99 $\%$ of the crack width. In other words:
\begin{equation}
\label{ep_99_def}
\frac{2\delta_3}{w}\geq 0.99, \quad x \in \left[(1-\varepsilon_{99})L,L \right].
\end{equation}
The temporal distribution of $\varepsilon_{99}$ is presented in Fig. \ref{d3_d99_500}b). It shows that in the early times of crack propagation the Newtonian layer, defined by condition \eqref{ep_99_def}, stretches over the entire crack length. Then the layer is reduced and reaches less than $1\%$ of the crack length in the final time instant.

\begin{figure}[htb!]
\begin{center}
\includegraphics[scale=0.5]{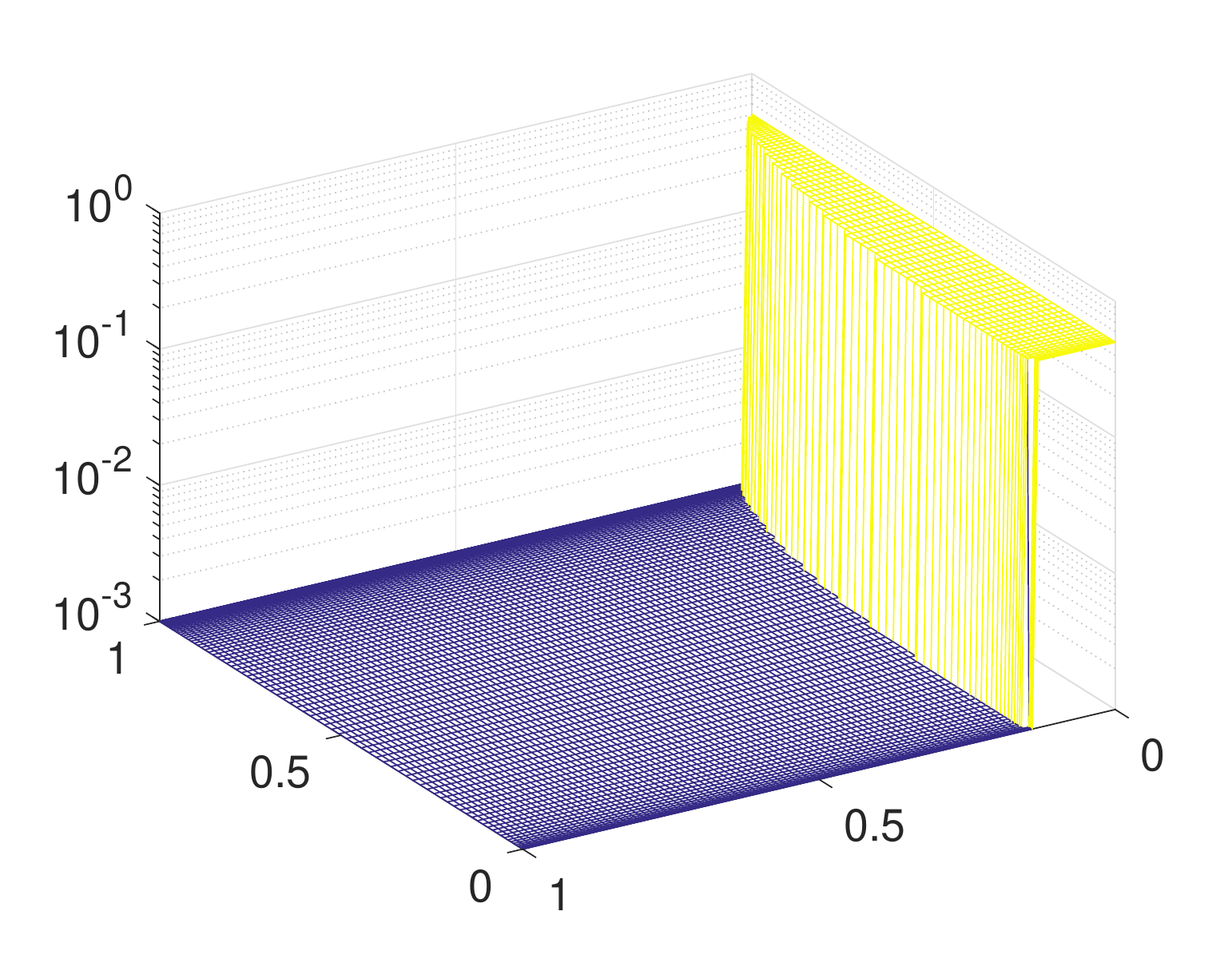}
\put(-75,10){$\frac{2y}{w}$}
\put(-185,20){$\frac{x}{L}$}
\put(-230,90){$\eta_\text{a}$}
\caption{Apparent viscosity distribution inside the fracture for the final time instant $t=20$ s, for $|\dot \gamma_2|=500.1$ s$^{-1}$.}
\label{eta_xy_500}
\end{center}
\end{figure}

We conclude this part of our analysis by presenting in Fig. \ref{eta_xy_500} the spatial distribution of the apparent viscosity $\eta_\text{a}$ over the fracture cross section in the final time instant ($t=20$ s). Only one of the symmetrical parts with respect to the $x$ axis is shown (i.e. $y \in[0,w/2]$). One can very clearly see the plateau of $\eta_0$ in the core of the flow (where $\dot \gamma$ assumes near-zero values). Simultaneously, over most of the fracture area the lower limiting viscosity $\eta_\infty$ is retained. Naturally, this zone increases when approaching the crack tip. It is only a very narrow strip between the aforementioned zones where the interim values of $\eta_\text{a}$ are achieved. These values correspond to the power-law section of the truncated power-law characteristics.

\subsection{$|\dot \gamma_2|=5\cdot 10^3$ $\text{s}^{-1}$}

\begin{figure}[htb!]
\begin{center}
\includegraphics[scale=0.5]{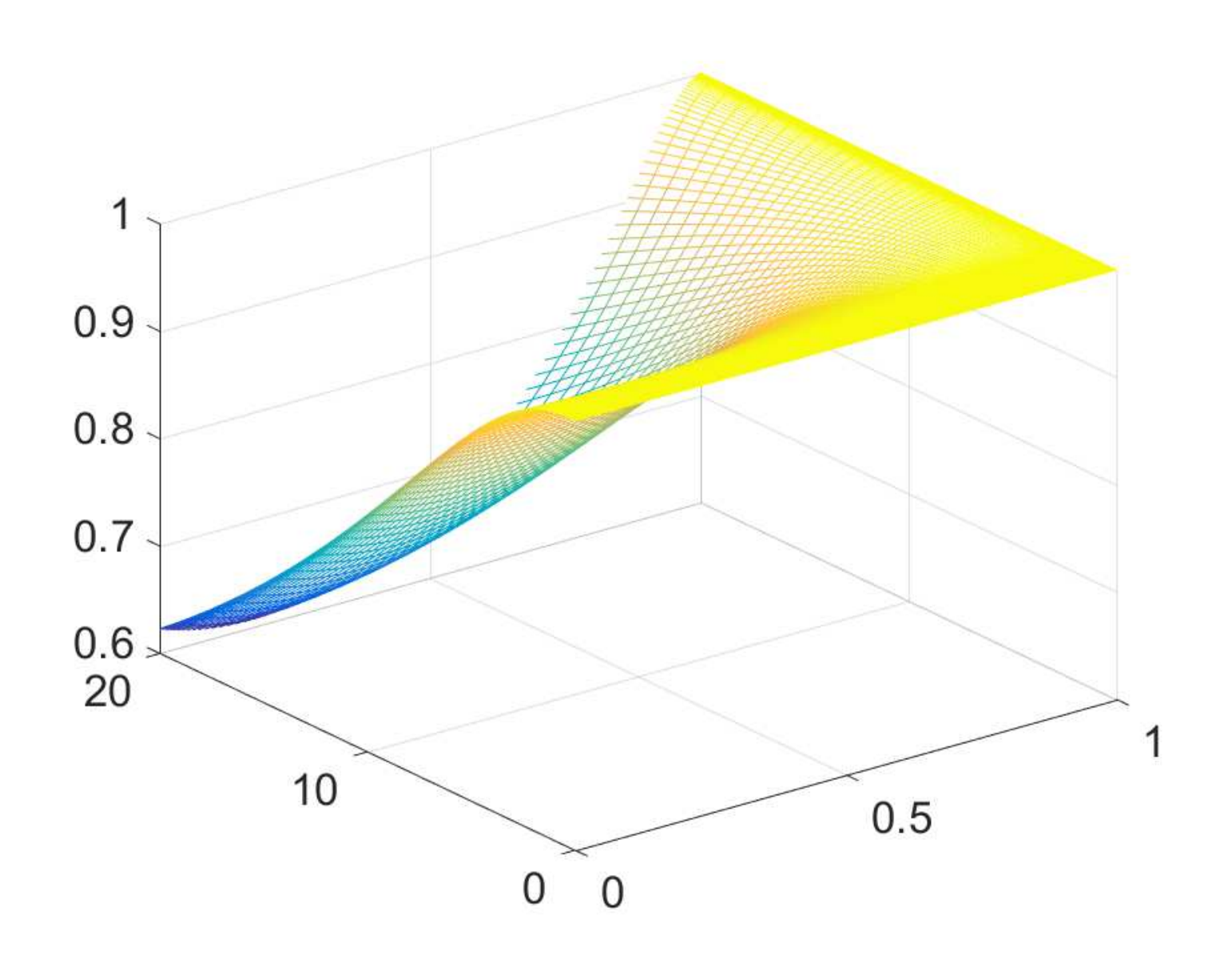}
\put(-60,10){$\frac{x}{L}$}
\put(-180,15){$t$}
\put(-225,95){$F$}
\includegraphics[scale=0.5]{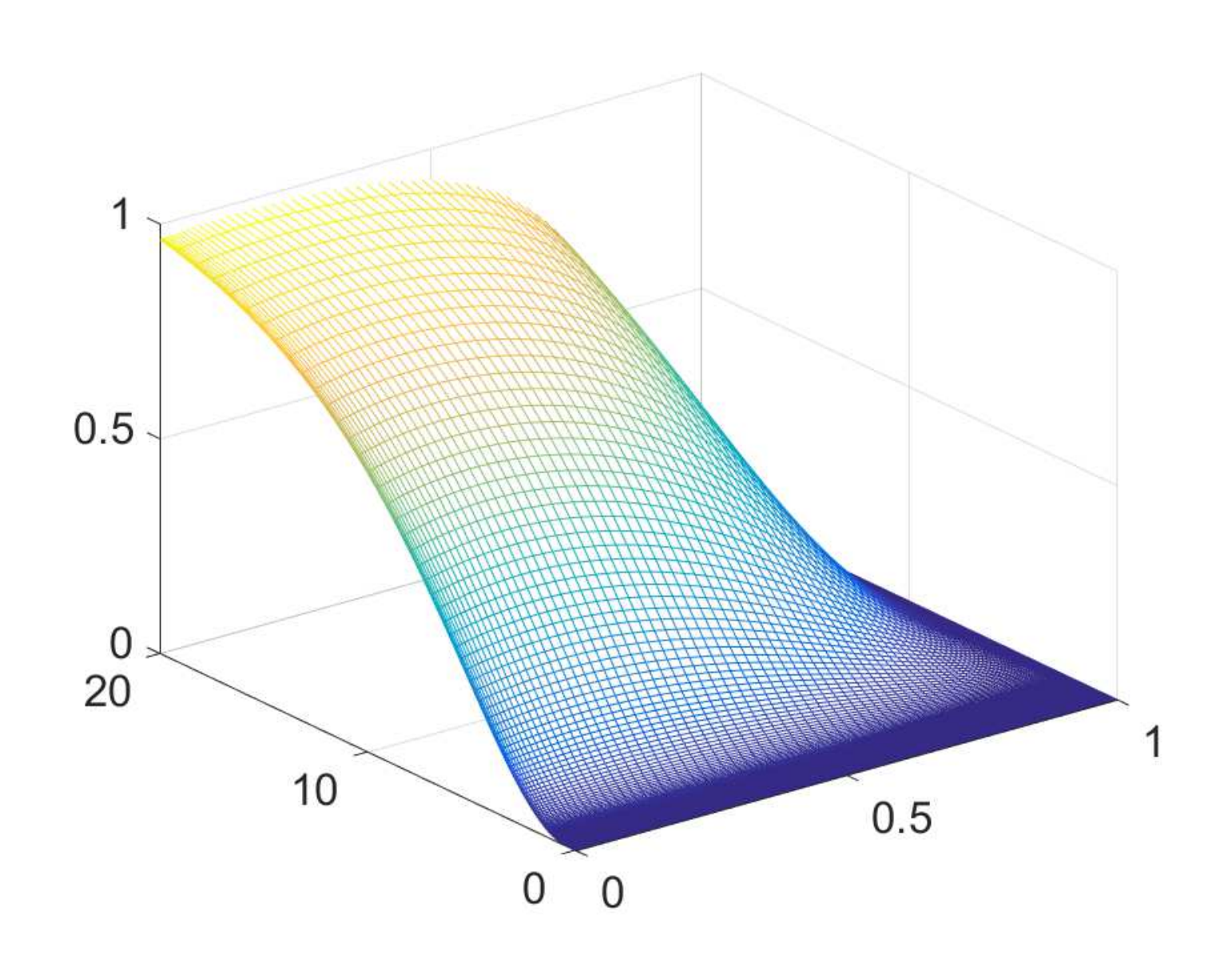}
\put(-60,10){$\frac{x}{L}$}
\put(-180,15){$t$}
\put(-227,95){$\frac{q}{q_\text{pl}}$}
\put(-446,165){$\textbf{a)}$}
\put(-220,165){$\textbf{b)}$}
\caption{a) The component function of fluid flow rate, $F(x,t)$. b) The ratio of the fluid flow rate, $q$, over the fluid flux computed according to the classical power-law solution, $q_\text{pl}$, for $|\dot \gamma_2|=5\cdot10^3$ s$^{-1}$.}
\label{F_5e3}
\end{center}
\end{figure}

The fluid flow rate component function, $F(x,t)$, for $|\dot \gamma_2|=5\cdot 10^3$ s$^{-1}$ is shown in Fig. \ref{F_5e3}a). This time $F$ is close to 1 over the entire crack length only in the initial times. With time growth it is only the near tip zone where the high shear rate Newtonian regime of flow is retained. On the other hand, the limiting value of $F$ for the low shear rate Newtonian regime ($\eta_\infty/\eta_0$) is not achieved at any point. This suggests that fluid flow regime evolves towards the one dominated by the power-law section of the viscosity characteristics. 

In order to substantiate this observation we present in Fig. \ref{F_5e3}b) a ratio of the computed fluid flow rate, $q$, over the fluid flux recreated according to the classical solution for the pure power-law fluid \citep{Perkowska_2016}:
\begin{equation}
\label{q_pl}
q_\text{pl}(x,t)=\frac{n}{2n+1}2^{-\frac{n+1}{n}}\left(-\frac{1}{C}\frac{\partial p}{\partial x}\right)^{1/n}w^\frac{2n+1}{n}.
\end{equation}
It can be seen in the figure that this ratio grows from zero at the initial times to almost unity over substantial part of the fracture length at the final time instant. Clearly, in the near-tip zone a zero value is retained due to the high shear rates dominance (i.e. the Newtonian model with viscosity $\eta_\infty$ holds exclusively).

\begin{figure}[htb!]
\begin{center}
\includegraphics[scale=0.5]{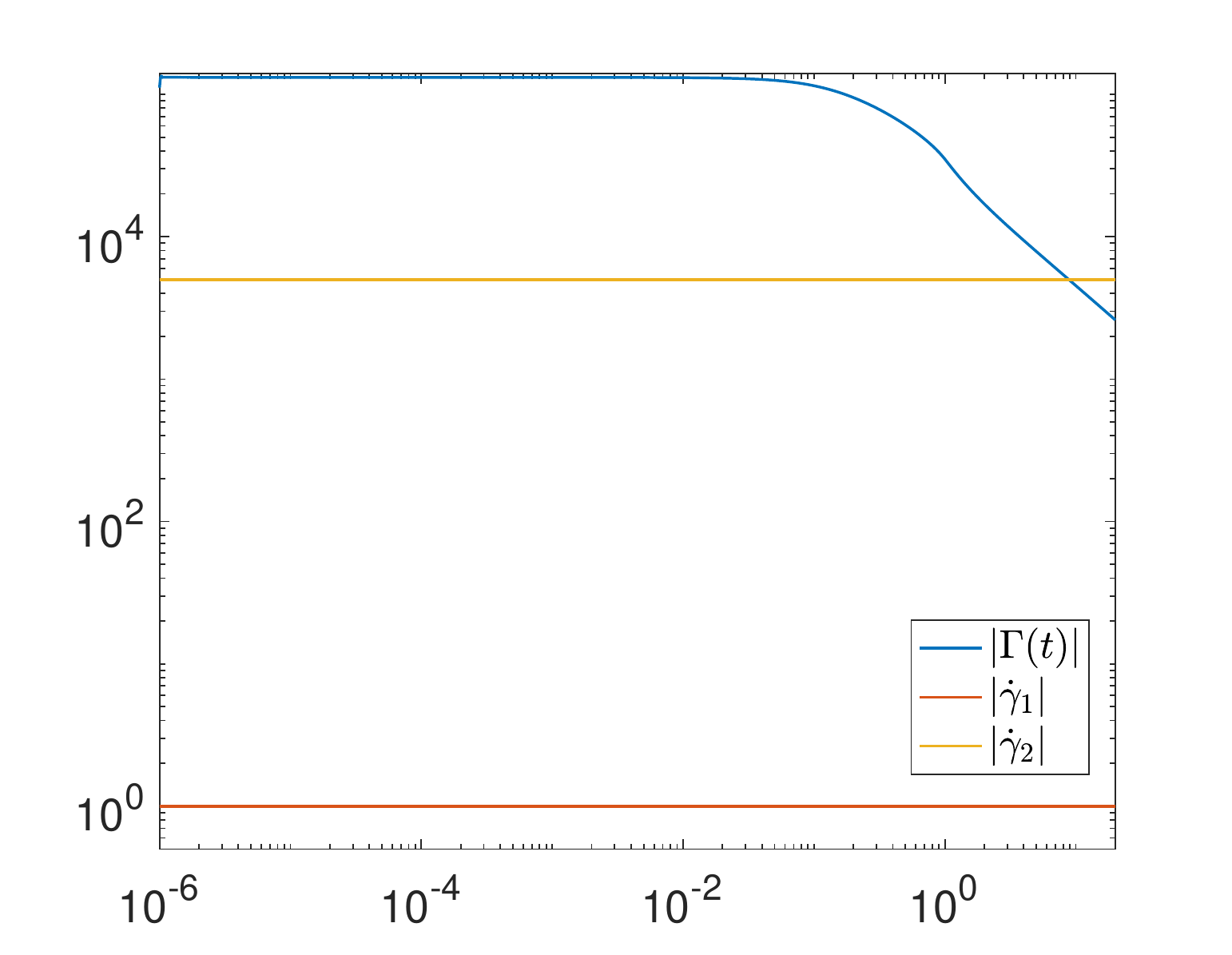}
\put(-120,-5){$t$}
\put(-230,80){$|\Gamma|$}
\caption{The fluid shear rate averaged over fracture surface, $|\Gamma(t)|$, for $|\dot \gamma_2|=5\cdot10^3$ s$^{-1}$.}
\label{Gam_5e3}
\end{center}
\end{figure}

The temporal evolution of the average shear rate is depicted in Fig. \ref{Gam_5e3}. Unlike the case of $|\dot \gamma_2|=500.1$ s$^{-1}$, this time $|\Gamma|$ falls below $|\dot \gamma_2|$ ($|\Gamma|=|\dot \gamma_2|$ at around $t=9$ s) to reach the shear rates range pertaining to the power-law section of viscosity characteristics.

\begin{figure}[htb!]
\begin{center}
\includegraphics[scale=0.5]{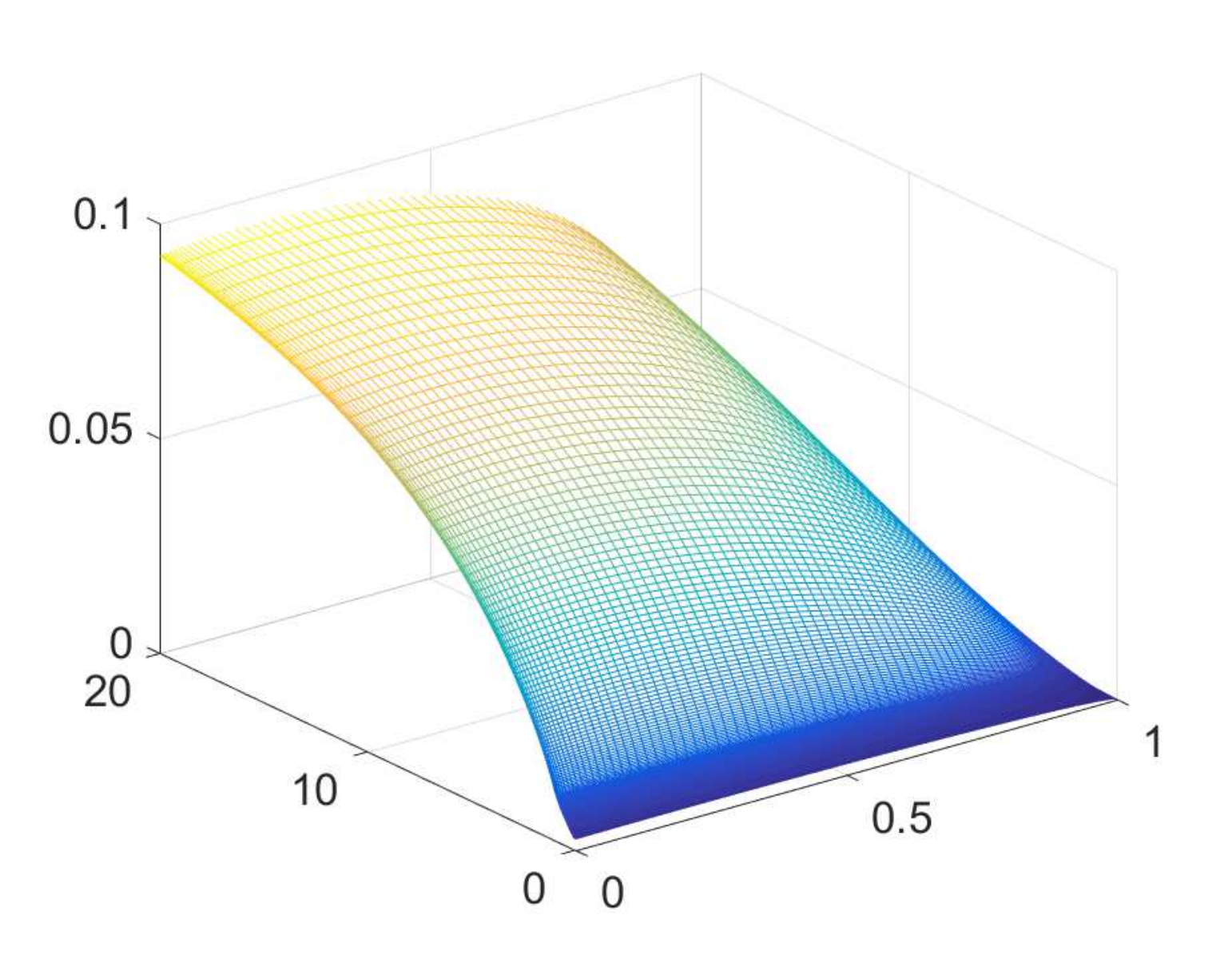}
\put(-60,10){$\frac{x}{L}$}
\put(-180,15){$t$}
\put(-230,95){$\frac{2\delta_1}{w}$}
\includegraphics[scale=0.5]{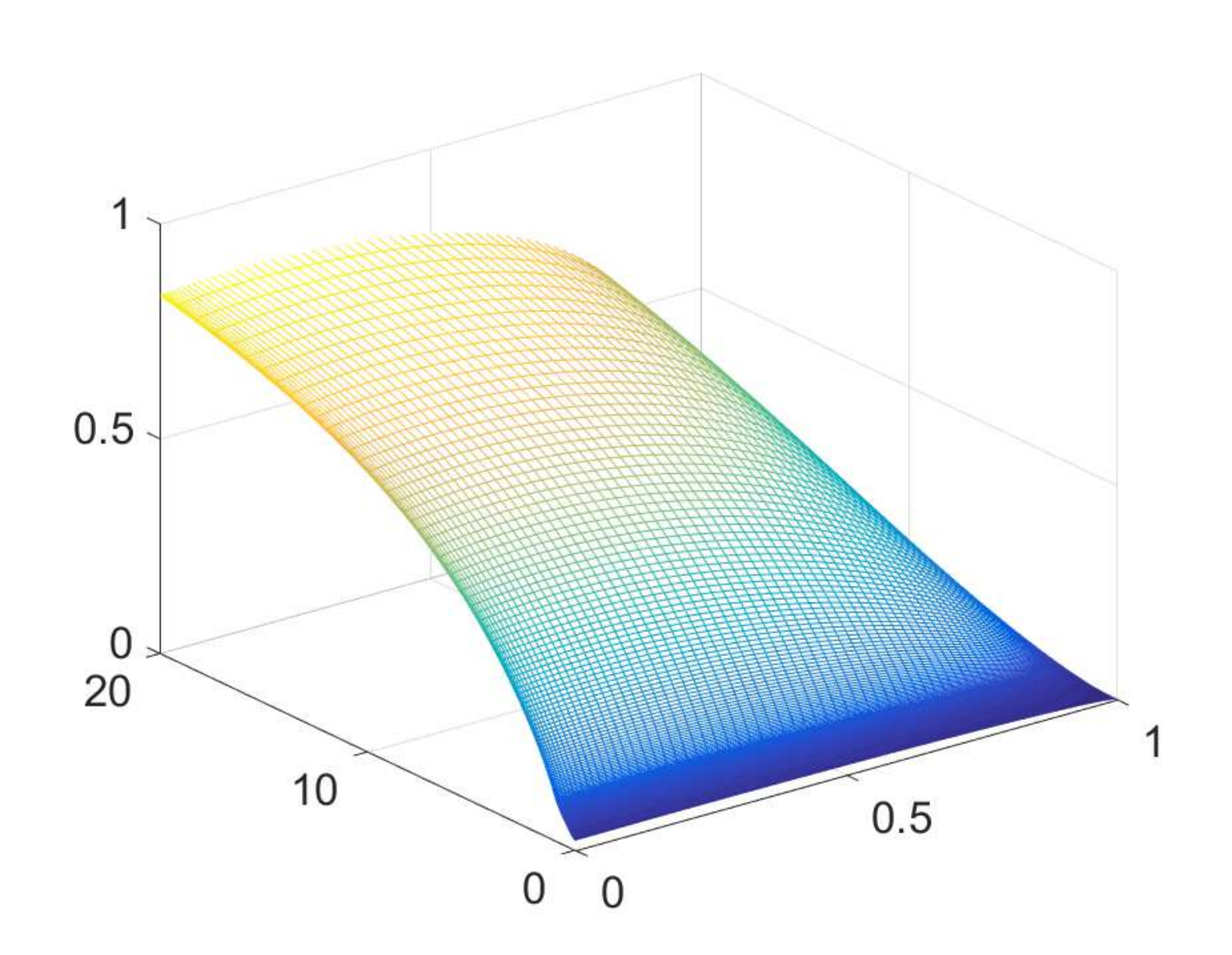}
\put(-60,10){$\frac{x}{L}$}
\put(-180,15){$t$}
\put(-224,95){$\frac{2\delta_2}{w}$}
\put(-446,165){$\textbf{a)}$}
\put(-220,165){$\textbf{b)}$}
\caption{The relative thickness of: a) the low shear rate layer, $\delta_1$, b) the intermediate shear rate layer, $\delta_2$, for $|\dot \gamma_2|=5\cdot10^3$ s$^{-1}$.}
\label{d1_d2_5e3}
\end{center}
\end{figure}

\begin{figure}[htb!]
\begin{center}
\includegraphics[scale=0.5]{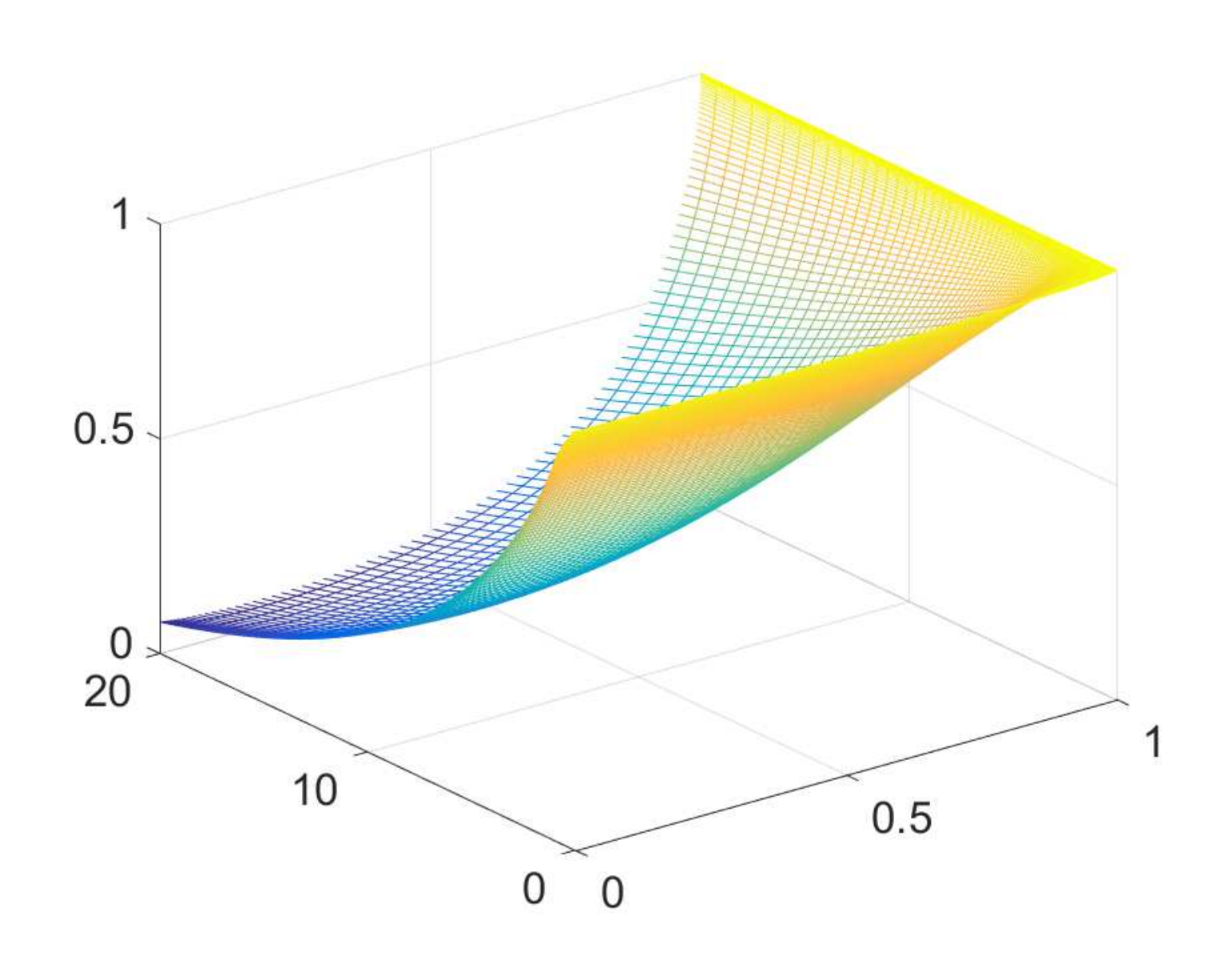}
\put(-60,10){$\frac{x}{L}$}
\put(-180,15){$t$}
\put(-230,95){$\frac{2\delta_3}{w}$}
\includegraphics[scale=0.5]{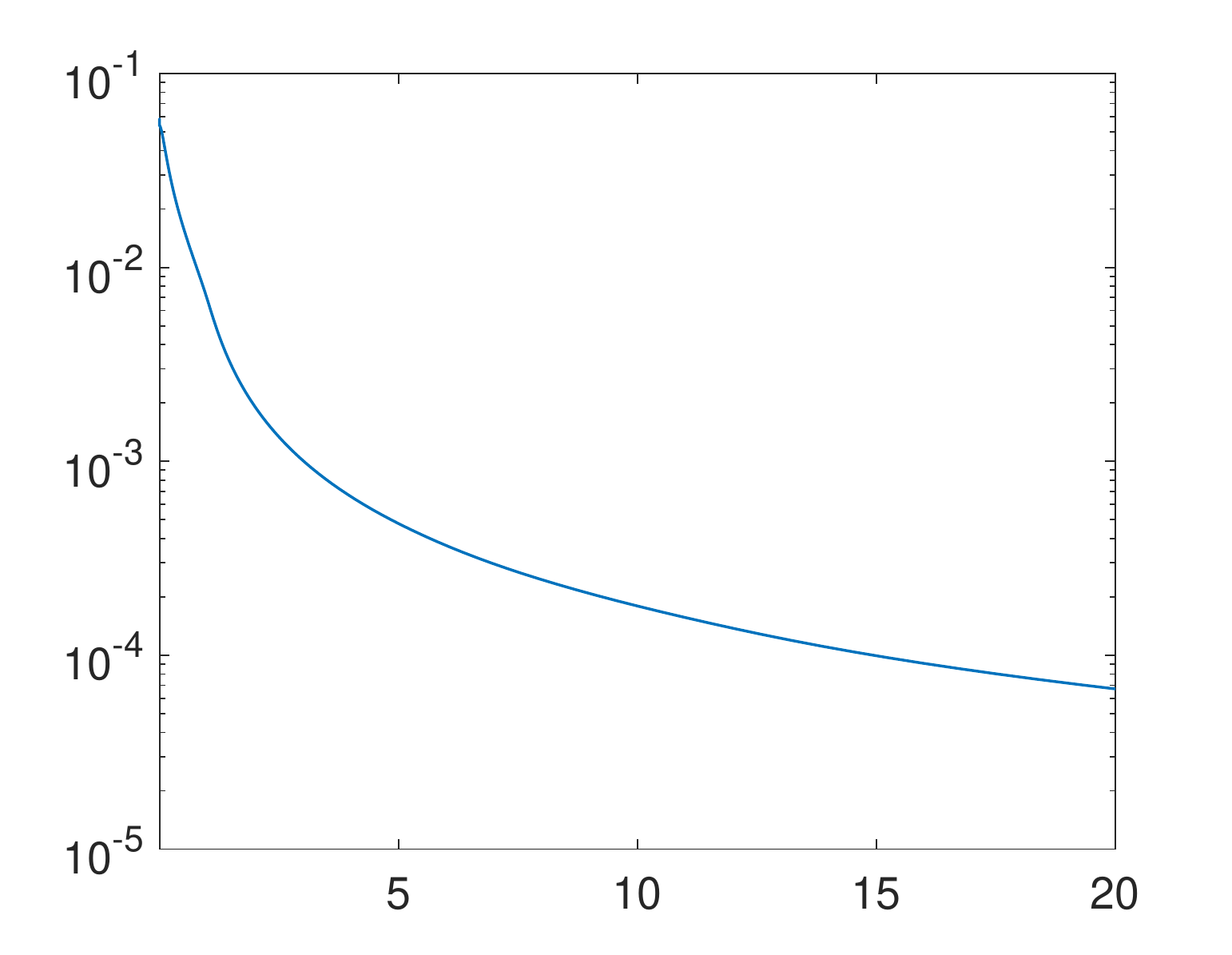}
\put(-110,-5){$t$}
\put(-446,165){$\textbf{a)}$}
\put(-220,165){$\textbf{b)}$}
\put(-225,90){$\varepsilon_{99}$}
\caption{a) The relative thickness of  the high shear rate layer. b) The relative length of the high shear rate Newtonian layer, $\varepsilon_{99}$, for $|\dot \gamma_2|=5\cdot10^3$ s$^{-1}$.}
\label{d3_d99_5e3}
\end{center}
\end{figure}

The relative thicknesses of respective shear rate layers are plotted in Figs. \ref{d1_d2_5e3}--\ref{d3_d99_5e3}a). As previously, the high shear rate Newtonian layer prevails at initial times and in the proximity of the crack tip (see Fig. \ref{d3_d99_5e3}a)). The relative contribution of low shear rate Newtonian layer has been reduced with respect to the case $|\dot \gamma_2|=500.1$ s$^{-1}$. However, this time one can clearly notice that it is the intermediate (power-law) layer that tends to occupy most of the fracture width at the later stages of crack propagation.

The relative distance over which the high shear rate Newtonian layer stretches along the crack length, $\varepsilon_{99}$, changes now from 5$\%$ at $t=0$ s to 0.07$\%$ at the final time instant (see Fig. \ref{d3_d99_5e3}b)).

\begin{figure}[htb!]
\begin{center}
\includegraphics[scale=0.5]{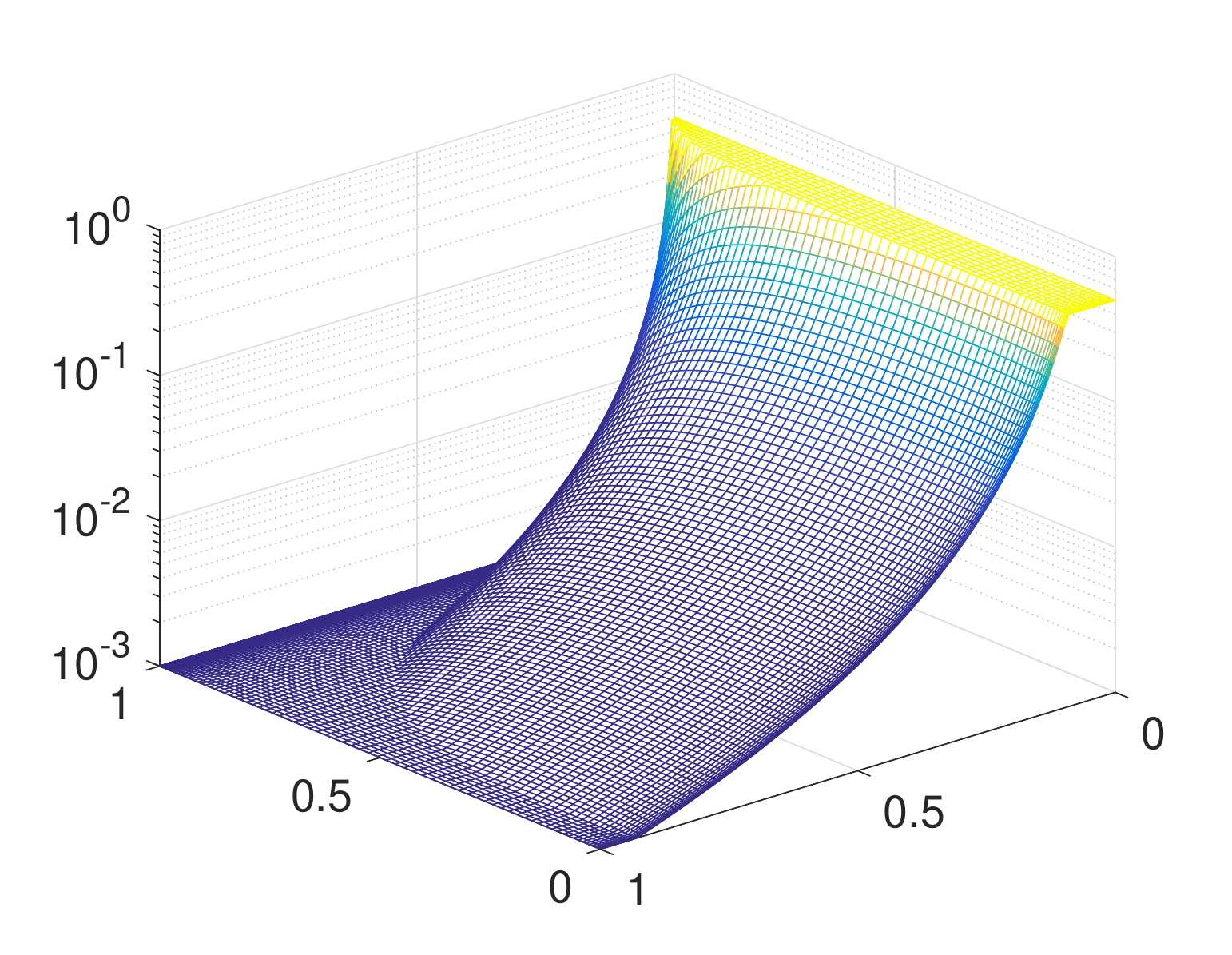}
\put(-70,10){$\frac{2y}{w}$}
\put(-180,20){$\frac{x}{L}$}
\put(-230,90){$\eta_\text{a}$}
\caption{Apparent viscosity distribution inside the fracture for the final time instant $t=20$ s, for $|\dot \gamma_2|=5\cdot10^3$ s$^{-1}$.}
\label{eta_xy_5e3}
\end{center}
\end{figure}

Finally, the distribution of the apparent viscosity $\eta_\text{a}$ over the fracture cross section at $t=20$ s (Fig. \ref{eta_xy_5e3}) becomes distinctly different from the one obtained for $|\dot \gamma_2|=500.1$ s$^{-1}$ (see Fig. \ref{eta_xy_500}). Even though the viscosity plateaus of $\eta_0$ and $\eta_\infty$ are again easily identified, the area over which the intermediate (power-law) characteristics holds has substantially increased.

\subsection{$|\dot \gamma_2|=10^6$ $\text{s}^{-1}$}

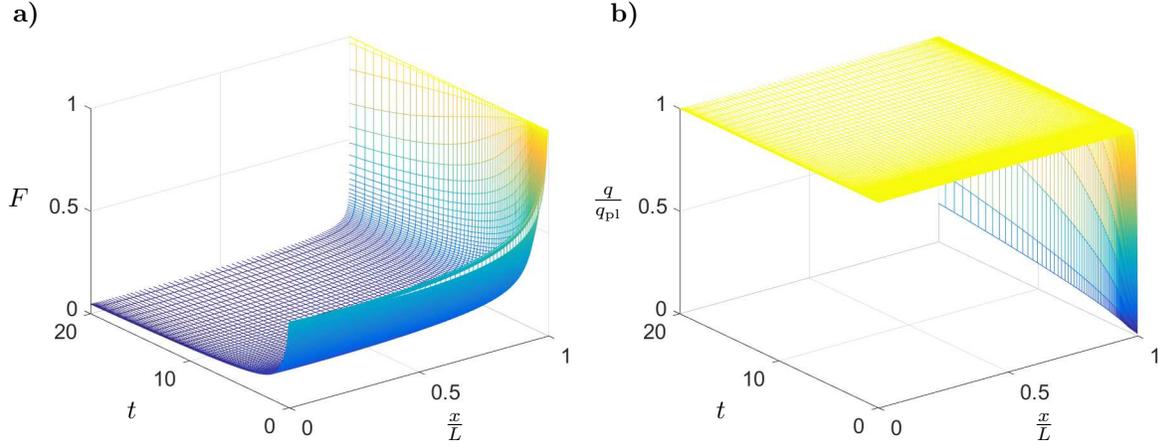
\begin{figure}[htb!]
\begin{center}
\includegraphics[scale=0.5]{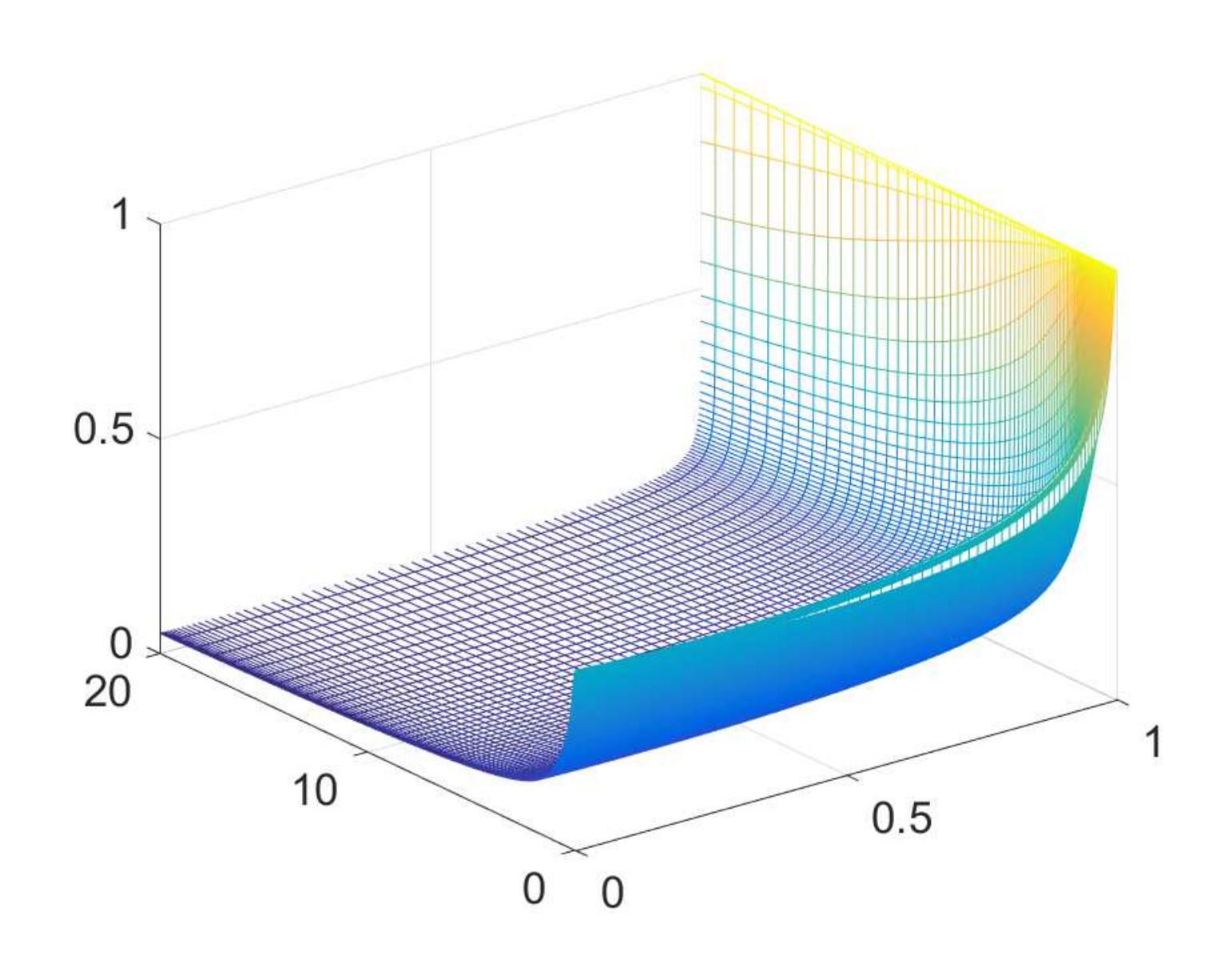}
\put(-60,10){$\frac{x}{L}$}
\put(-180,15){$t$}
\put(-225,95){$F$}
\includegraphics[scale=0.5]{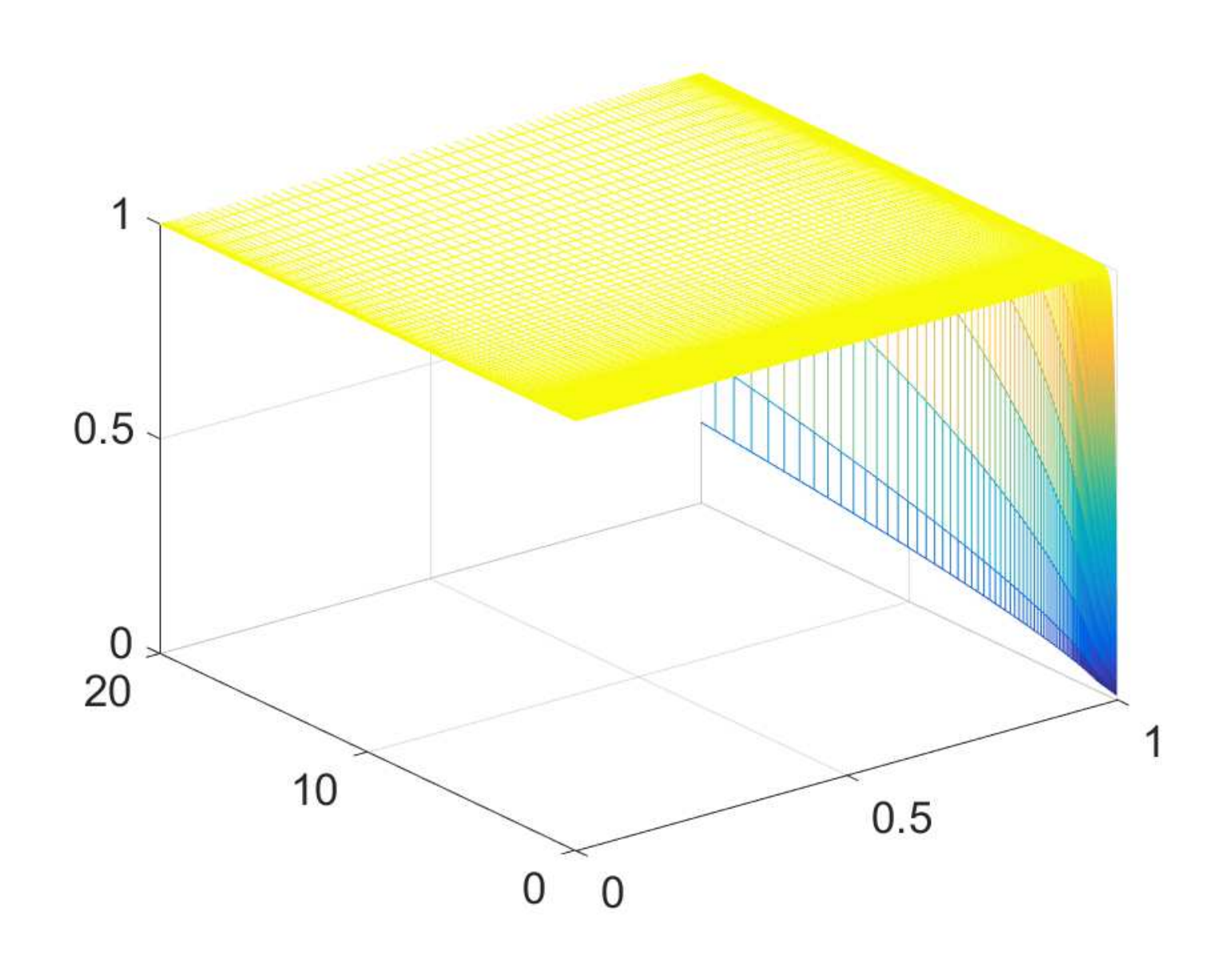}
\put(-60,10){$\frac{x}{L}$}
\put(-180,15){$t$}
\put(-227,95){$\frac{q}{q_\text{pl}}$}
\put(-446,165){$\textbf{a)}$}
\put(-220,165){$\textbf{b)}$}
\caption{a) The component function of fluid flow rate, $F(x,t)$. b) The ratio of the fluid flow rate, $q$, over the fluid flux computed according to the classical power-law solution, $q_\text{pl}$, for $|\dot \gamma_2|=10^6$ s$^{-1}$.}
\label{F_1e6}
\end{center}
\end{figure}

In the last analyzed in detail case, the fluid flux component function, $F$, tends to unity only in the near-tip zone regardless of the stage of crack propagation (Fig. \ref{F_1e6}a)). Simultaneously, the fluid flow rate computed with the pure power-law viscosity, $q_\text{pl}$, mimics perfectly the actual flux, $q$, everywhere except in the immediate vicinity of the fracture tip (Fig. \ref{F_1e6}b)).

\begin{figure}[htb!]
\begin{center}
\includegraphics[scale=0.5]{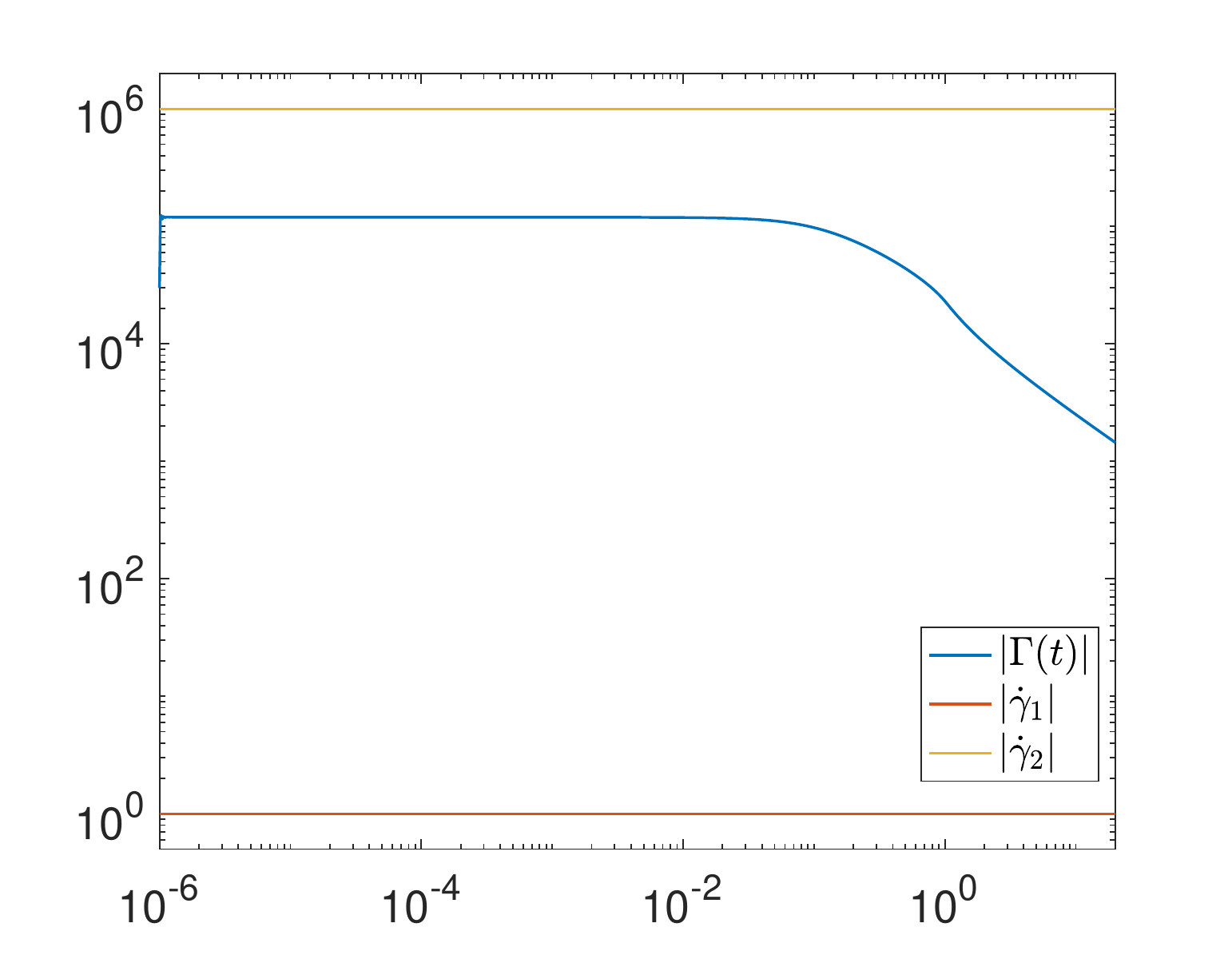}
\put(-120,-5){$t$}
\put(-230,80){$|\Gamma|$}
\caption{The fluid shear rate averaged over the fracture surface, $|\Gamma(t)|$, for $|\dot \gamma_2|=10^6$ s$^{-1}$.}
\label{Gam_1e6}
\end{center}
\end{figure}

An explanation of this fact becomes obvious when one looks at the graph of the average shear rate $\Gamma(t)$ (Fig. \ref{Gam_1e6}). It shows that over the entire temporal domain the following relation is satisfied:
\[
|\dot \gamma_1|<|\Gamma|<|\dot \gamma_2|.
\]

\begin{figure}[htb!]
\begin{center}
\includegraphics[scale=0.5]{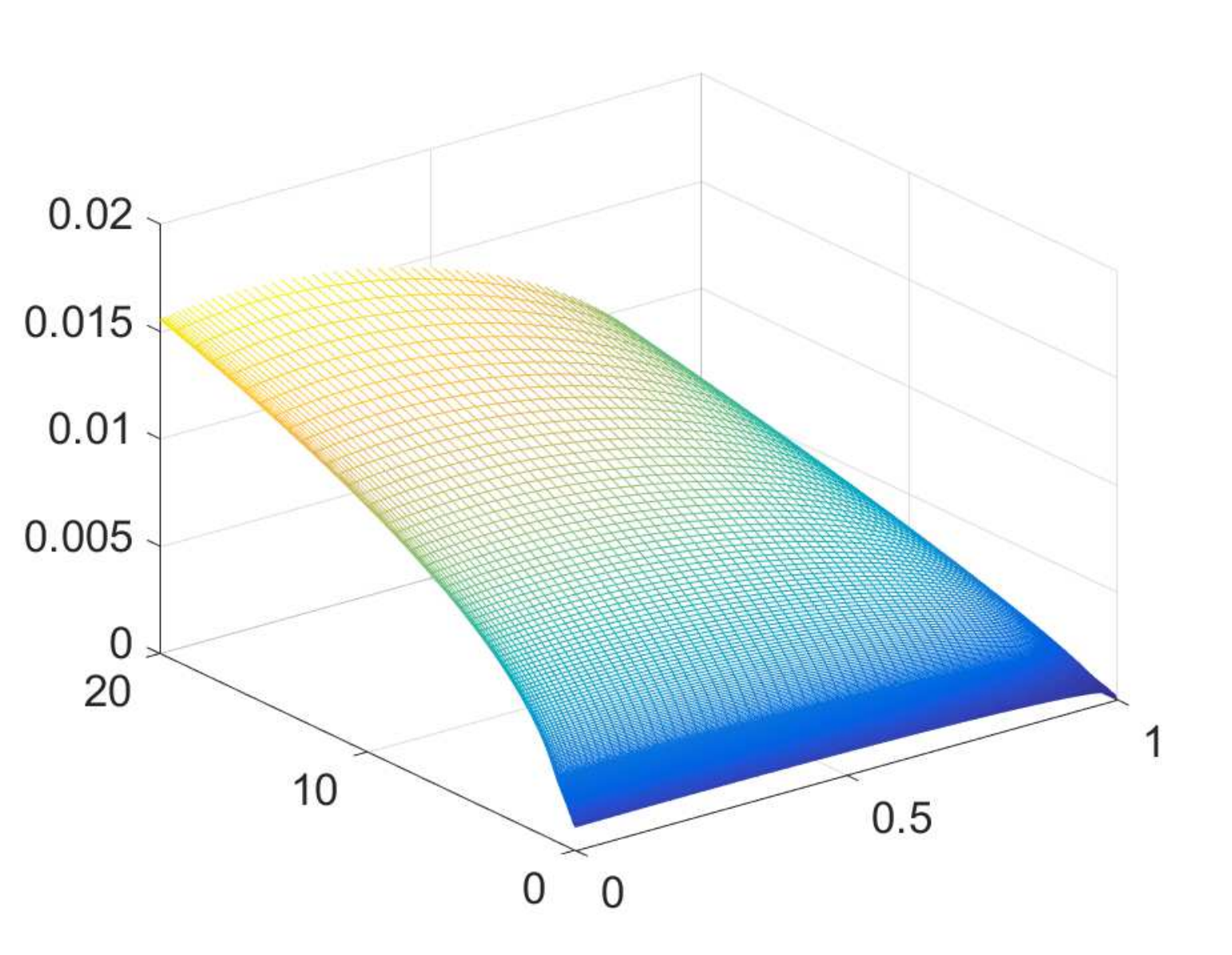}
\put(-60,10){$\frac{x}{L}$}
\put(-180,15){$t$}
\put(-230,95){$\frac{2\delta_1}{w}$}
\includegraphics[scale=0.5]{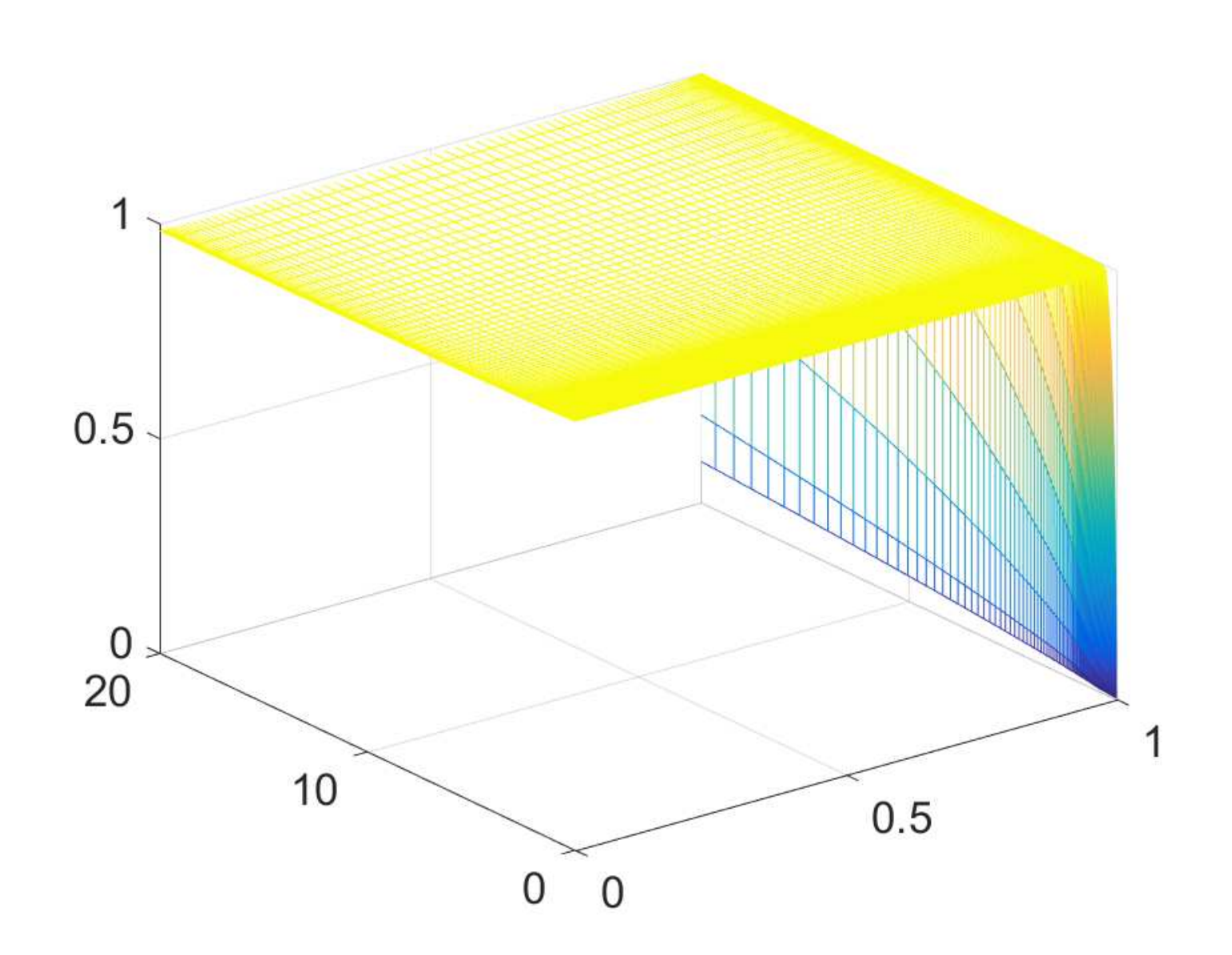}
\put(-60,10){$\frac{x}{L}$}
\put(-180,15){$t$}
\put(-223,95){$\frac{2\delta_2}{w}$}
\put(-446,165){$\textbf{a)}$}
\put(-220,165){$\textbf{b)}$}
\caption{The relative thickness of: a) the low shear rate layer, $\delta_1$, and b) the intermediate shear rate layer, $\delta_2$, for $|\dot \gamma_2|=10^6$ s$^{-1}$.}
\label{d1_d2_1e6}
\end{center}
\end{figure}

\begin{figure}[htb!]
\begin{center}
\includegraphics[scale=0.5]{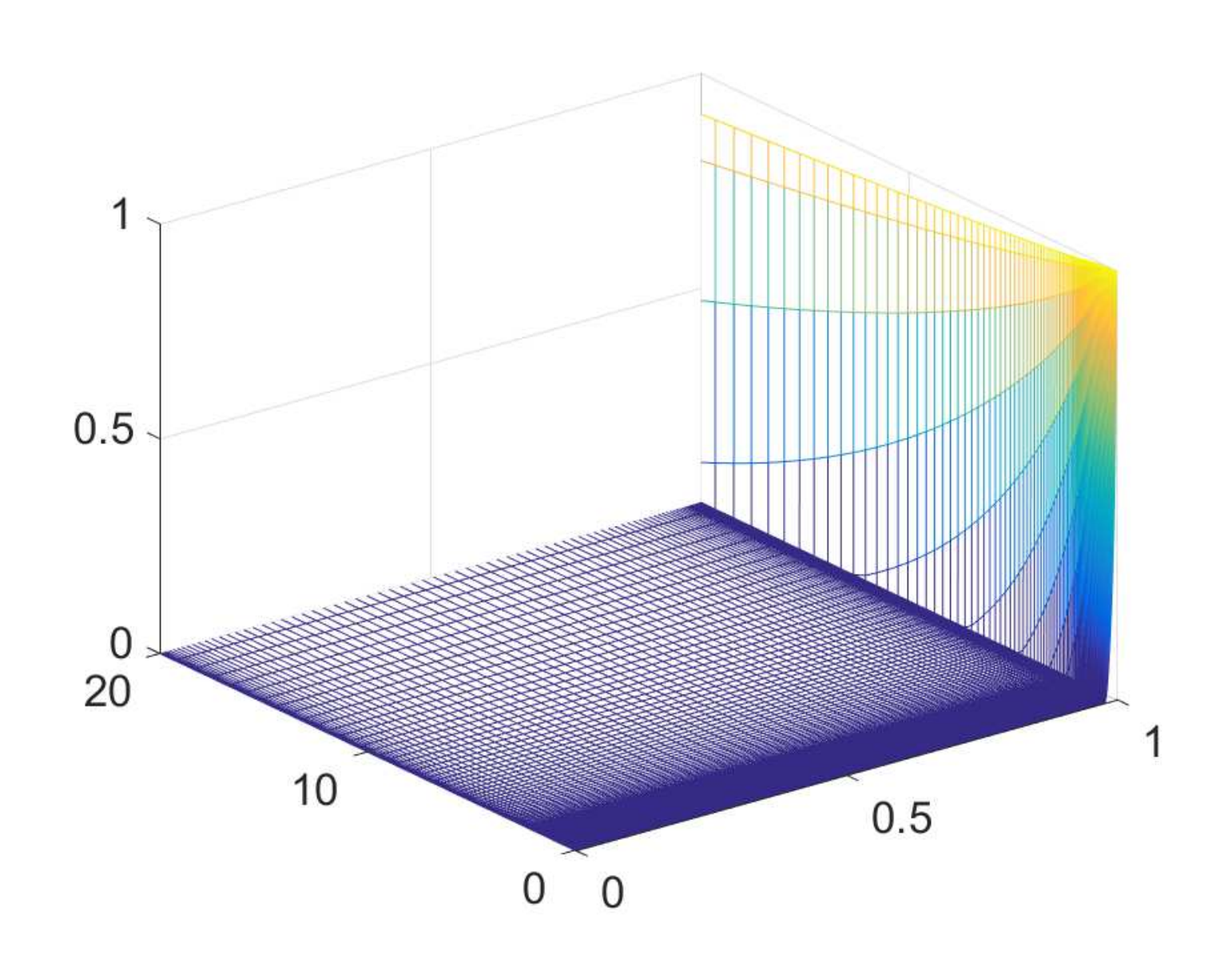}
\put(-60,10){$\frac{x}{L}$}
\put(-180,15){$t$}
\put(-230,95){$\frac{2\delta_3}{w}$}
\includegraphics[scale=0.5]{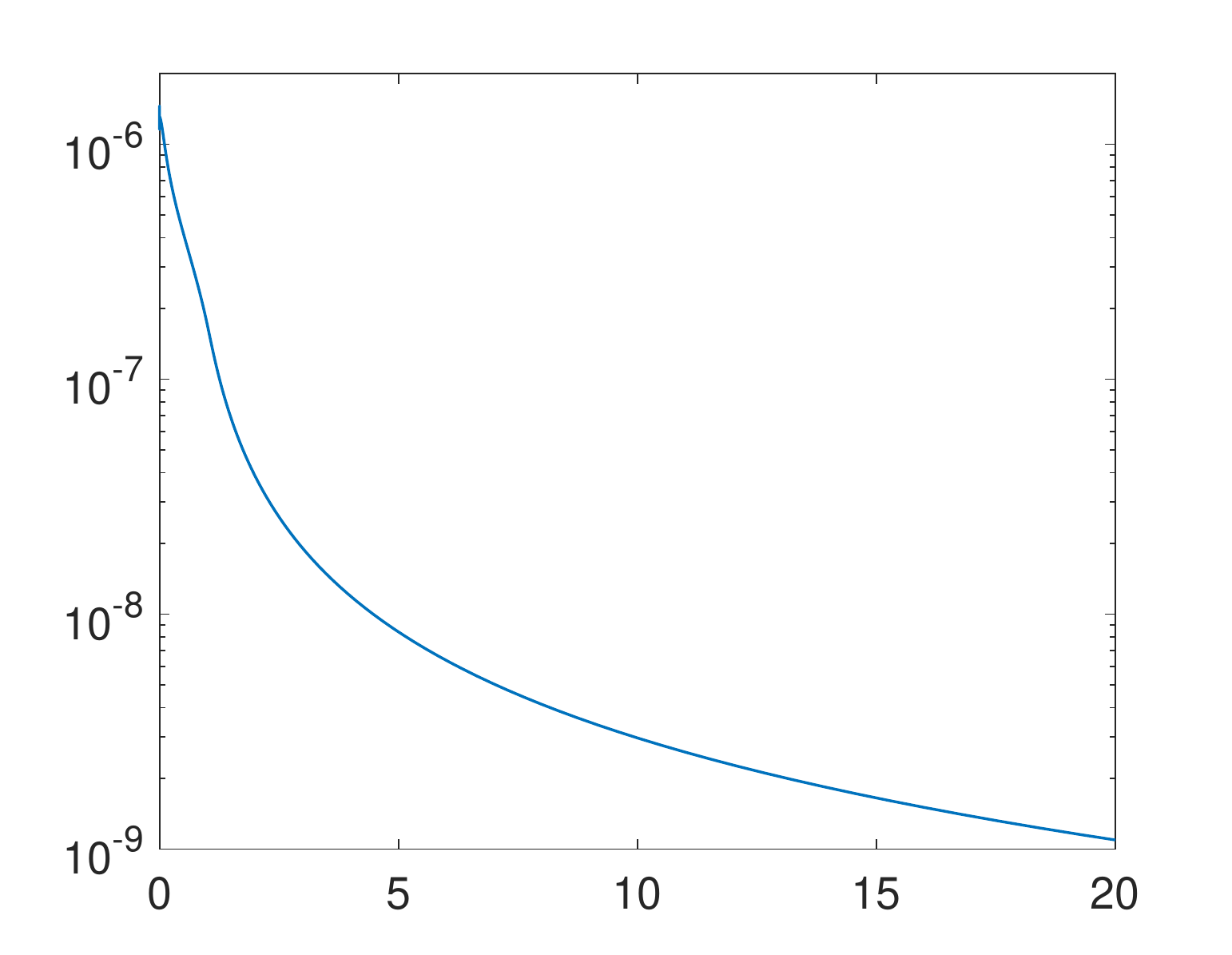}
\put(-110,-5){$t$}
\put(-446,165){$\textbf{a)}$}
\put(-220,165){$\textbf{b)}$}
\put(-225,90){$\varepsilon_{99}$}
\caption{a) The relative thickness of  the high shear rate layer. b) The relative length of the high shear rate Newtonian layer, $\varepsilon_{99}$, for $|\dot \gamma_2|=10^6$ s$^{-1}$.}
\label{d3_d99_1e6}
\end{center}
\end{figure}
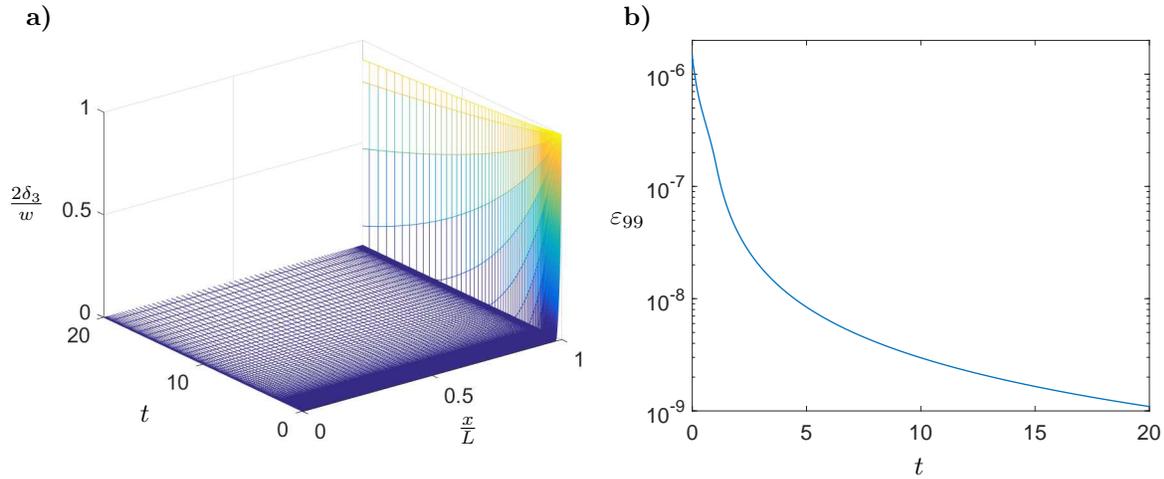

Also the distribution of the relative thicknesses of respective shear rate layers (Figs. \ref{d1_d2_1e6}--\ref{d3_d99_1e6}a)) reveals that the intermediate one (the power-law layer of thickness $\delta_2$) occupies almost the whole width of the fracture. In comparison with the case of $|\dot \gamma_2|=500.1$ s$^{-1}$, the relative thickness of the low shear rate Newtonian layer ($\eta_\text{a}=\eta_0$) has been reduced by approximately ten times, while the relative thickness of the high shear rate layer ($\eta_\text{a}=\eta_\infty$) has shrunk to zero everywhere outside the near-tip zone. The relative coverage of the latter along the crack length, in terms of $\varepsilon_{99}$, ranges from $10^{-4}\%$ at the initial time to $10^{-7}\%$ at $t=20$ s (see Fig. \ref{d3_d99_1e6}b)).

\begin{figure}[htb!]
\begin{center}
\includegraphics[scale=0.5]{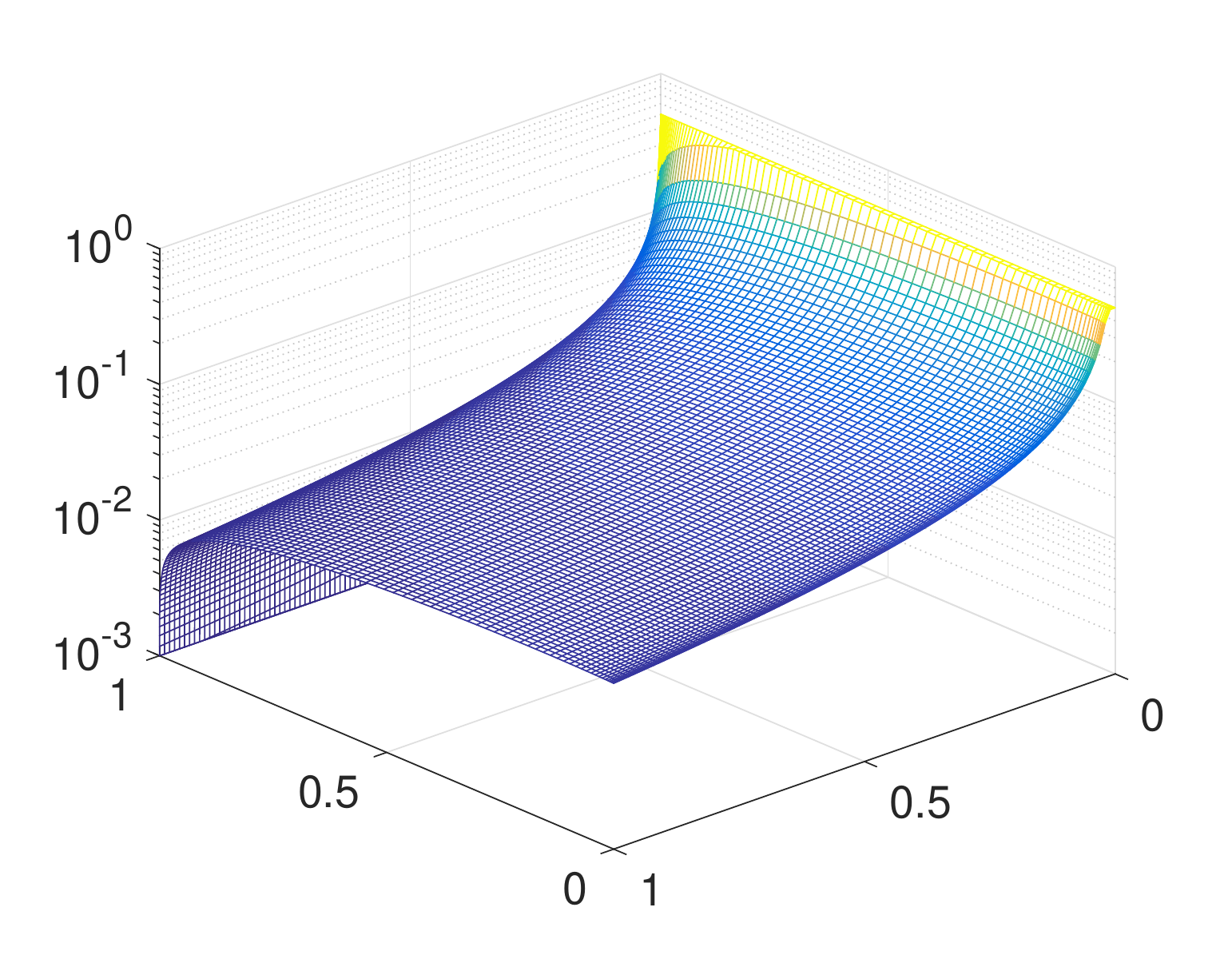}
\put(-70,10){$\frac{2y}{w}$}
\put(-180,20){$\frac{x}{L}$}
\put(-230,90){$\eta_\text{a}$}
\caption{Apparent viscosity distribution inside the fracture for the final time instant $t=20$ s, for $|\dot \gamma_2|=10^6$ s$^{-1}$.}
\label{eta_xy_1e6}
\end{center}
\end{figure}

The spatial distribution of apparent viscosity over the fracture cross section at $t=20$ s is shown in Fig. \ref{eta_xy_1e6}. Two crucial differences can be noticed with respect to the previous cases (see Fig. \ref{eta_xy_500} and Fig. \ref{eta_xy_5e3}). Firstly, the Newtonian plateau of $\eta_0$ has disappeared almost completely. Secondly, there is no more any distinct Newtonian plateau of $\eta_\infty$. The limiting viscosity  $\eta_\infty$ is retained virtually only at the crack tip. Instead, over almost the entire fracture footprint the apparent viscosity values correspond to the intermediate (power-law) range of the truncated power-law characteristics.

\section{Discussion of the results}
\label{dis}

The general results presented in the first part of this analysis confirm that for varying $|\dot \gamma_2|$ the solution evolves in a parametric space encompassed by two limiting solutions obtained for Newtonian fluids, one  for viscosity $\eta_0$ and the other for $\eta_\infty$. For $|\dot \gamma_2|\leq10^3$ s$^{-1}$ the results are indistinguishable from those pertaining to  $\eta_\infty$. Obviously, the aforementioned value of $|\dot \gamma_2|$ is not a strict limit and one can expect that even above  this magnitude the resulting data can coincide well with the limiting Newtonian solution. Clearly, this could be changed by increasing $|\dot \gamma_1|$. However, the accepted value is approximately the upper limit for fracturing fluids \citep{Wrobel_2020,Lavrov_2015,Lecampion_2018,Habibpour_2017}. For $|\dot \gamma_1|<1$ s$^{-1}$ the described situation would remain unchanged. On the other hand, the second limiting Newtonian solution with $\eta_0$ is not reached even for $|\dot \gamma_2|=10^8$ s$^{-1}$, which suggests that rather unrealistic values of $|\dot \gamma_2|$ are needed to attain the low-shear rate limit. 

The aforementioned evolution of results between the limiting regimes is caused by the interplay between the predefined fluid rheology, the pumping rate and the resulting fracture geometry, when material parameters of solid are assumed to be constant. In order to illustrate the underlying mechanisms three cases of $|\dot \gamma_2|$ have been analyzed in detail. 

The first one, $|\dot \gamma_2|=500.1$ s$^{-1}$, constitutes an example of the high shear rate regime. Here, the average values of shear rates, $|\Gamma|$ are higher than $|\dot \gamma_2|$ throughout the whole duration of fracture extension. The  high shear rate layer of thickness $\delta_3$ tends to occupy almost the whole width of the fracture, with its relative coverage, $\varepsilon_{99}$, reaching 100 $\%$ of the crack length at the initial stages of crack propagation. Even though the high shear rate Newtonian layer shrinks with time, the results remain in close coincidence with those for the Newtonian fluid with $\eta_\infty$ up to the end of simulations. 

The second example, $|\dot \gamma_2|=5\cdot 10^3$ s$^{-1}$, demonstrates a case where a gradual transition from the high shear rate regime to a mixed mode takes place. This time $|\Gamma|$ intersects at some point  the level of  $|\dot \gamma_2|$  and the power-law section of viscosity characteristics becomes dominant. Notably, the intermediate shear rate layer of thickness $\delta_2$ grows with time primarily at the expense of the high shear rate layer ($\delta_3$).  The spatial coverage of the latter, $\varepsilon_{99}$, is of one order of magnitude lower than it was for $|\dot \gamma_2|=500.1$ s$^{-1}$.

Finally, the third example, $|\dot \gamma_2|= 10^6$ s$^{-1}$, illustrates a situation where the power-law section of viscosity characteristics dominates the process throughout its whole duration. The average values of the shear rates, $|\Gamma|$, are now well inside the interval defined by $|\dot \gamma_1|$ and $|\dot \gamma_2|$. It is only in the immediate vicinity of the crack tip where the high shear rate Newtonian layer is predominant. The spatial coverage of this zone, $\varepsilon$, however, becomes extremely small. The resulting fluid flow rate is in a very good agreement with the one recreated by the classical solution for the power-law fluid. Thus, the latter model can be in this case a credible substitute for the truncated power-law. The last conclusion can be also confirmed by results obtained in \cite{Wrobel_2020} for the HPG fluid in the framework of the PKN HF problem.

It should be noted that, if the fluid lag is accounted for, the near tip high shear rate Newtonian zone can be reduced or may not exist at all. This issue becomes especially important in those cases where the lag between the crack tip and the fracture front is relatively large (e.g. small solid toughness and small tip underpressure - see  \cite{Garagash_2006a}). 

The description of the above three cases of $|\dot \gamma_2|$ enables one to understand the evolution of the crack propagation regimes between the two limiting Newtonian modes, i.e. these examples illustrate the nature and direction of the basic trends.

\section{Final conclusions}
\label{con}

In this paper a problem of a hydraulic fracture driven by shear-thinning fluid is analyzed. The rheological properties of the fracturing fluid are described by the four parameter truncated power-law model. The KGD (plane strain) fracture geometry is assumed in a modified formulation that accounts for the hydraulically induced tangential tractions on the crack surfaces. For some typical parameters of the HF process a number of simulations are performed with varying magnitude of the limiting cut-off shear rates. The underlying mechanisms that govern the fracture evolution between limiting regimes are investigated.  

The following conclusions can be drawn from the conducted analysis:
\begin{itemize}
\item{The rheological properties of fracturing fluids affect crucially the process of hydraulic fracture not only by the limiting values of viscosity, but also by the range of fluid shear rates over which variation of viscosity occurs. }
\item{When neglecting the fluid lag, it is always the near-tip zone where the Newtonian model of flow with viscosity $\eta_\infty$ holds. This is due to high fluid shear rates caused by negative singular pressure gradient. This conclusion holds for any fracturing fluid that exhibits high shear rate cut-off viscosity.}
\item{For fixed parameters of the HF process the size of the Newtonian high shear rate zone depends on the value of cut-off shear rate $|\dot \gamma_2|$ and the stage of crack propagation. In general, the spatial coverage of this zone decreases as the fracture evolves (with the crack length growth and the velocity decrease).}
\item{With growing $|\dot \gamma_2|$ the fluid flow inside the fracture evolves from the high shear rate Newtonian regime towards the intermediate mode governed by power-law section of the viscosity characteristics. Locally, the high shear rate regime is always retained at the crack tip (provided that there is no lag). It seems that transition to the low shear rate Newtonian regime of flow requires either unrealistically high values of $|\dot \gamma_1|$  or very long times of the process.}
\item{As the truncated power-law rheology has been recognized to mimic well the Carreau model \citep{Wrobel_2020}, the results of this analysis can be extended to more sophisticated four parameter rheological models of fluid that exhibit limiting cut-off viscosities.}
\item{Methodology and computational schemes presented in this paper and previous publication by \cite{Wrobel_2020} can be used to provide feedback on the expected shear rate values in the HF treatments needed for proper execution of the rheometric measurements of fracturing fluids. }
\item{Methodology and computational schemes introduced in this paper and the previous publication by \cite{Wrobel_2020} can by used together with proper scaling of the problem to define a parametric space of applicability for the power-law model when the latter is considered  an approximation of a certain rheological law from the category of the generalized Newtonian fluids. }
\item{I the case of rough walled fractures the channelization effects related to the truncated power-law rheology \citep{Felisa_2018,Lavrov_2013} can affect significantly the local flow patterns especially in the proximity of the boundaries between respective shear rate layers over the fracture footprint.  This can also make questionable the validity of the lubrication theory when pronounced asperities of the fracture faces are present. }
\end{itemize}

\section*{Acknowledgments}
The authors are thankful to Dr Monika Perkowska and Dr Martin Dutko for their useful comments and discussions. GM acknowledges the Royal Society for the Wolfson Research Merit Award and Welsh Government for the Ser Cymru Future Generations Industrial Fellowship.
\noindent

\vspace{5mm}
\noindent
{\bf Funding:} 
This work was funded by European Regional Development Fund and the Republic of Cyprus
through the Research Promotion Foundation (RESTART 2016 - 2020 PROGRAMMES, Excellence Hubs,
Project EXCELLENCE/1216/0481).

\end{document}